\documentclass[final,3p,12pt]{elsarticle}

\usepackage{amsmath}
\usepackage{amssymb}
\usepackage{amsfonts}
\usepackage{xcolor}
\usepackage{adjustbox}
\usepackage{array}
\usepackage[colorlinks=true,linkcolor=blue,urlcolor=blue,citecolor=blue]{hyperref}
\usepackage{times}
\usepackage[normalem]{ulem}

\journal{Nucl.~Phys.~B}

\usepackage{tikz}
\usetikzlibrary{arrows,shapes,positioning}
\usetikzlibrary{decorations.markings}
\tikzstyle arrowstyle=[scale=1]
\tikzstyle directed=[postaction={decorate,decoration={markings,
    mark=at position .65 with {\arrow[arrowstyle]{stealth}}}}]
\tikzstyle endreversedirected=[postaction={decorate,decoration={markings,
    mark=at position 1.0 with {\arrow[arrowstyle]{stealth}}}}]
\tikzstyle enddirected=[postaction={decorate,decoration={markings,
    mark=at position 1.0 with {\arrow[arrowstyle]{stealth}}}}]
\tikzstyle reverse directed=[postaction={decorate,decoration={markings,
    mark=at position .65 with {\arrowreversed[arrowstyle]{stealth};}}}]
\newcommand{\kaysubsection}[1]{\section{#1}}

\newcommand{\Mathematica}[1]{}

\newcommand{\suppress}[1]{}
\newcommand{\Eq}[1]{Eq.~(\ref{#1})}
\newcommand{\eq}[1]{(\ref{#1})}

\newcommand{\half}{\frac12}
\newcommand{\bea}{\begin{eqnarray}}
\newcommand{\eea}{\end{eqnarray}}
\newcommand{\beq}{\begin{equation}}
\newcommand{\eeq}{\end{equation}}
\newcommand{\be}{\begin{equation}}
\newcommand{\ee}{\end{equation}}

\newcommand{\rme}{\mathrm{e}}
\newcommand{\rmd}{\mathrm{d}}
\newcommand{\nn}{\nonumber}
\newcommand{\tr}{\mathrm{tr}}
\renewcommand{\epsilon}{\varepsilon}
\newcommand{\nott}[1]{}

\newcommand{\ffig}[2]{\includegraphics[width=#1]{./#2}}

\newlength{\bilderlength}
\newcommand{\bilderscale}{0.35}
\newcommand{\textbilderscale}{0.25}
\newcommand{\usebilderscale}{\bilderscale}
\newcommand{\bilderskip}{\hspace*{0.8ex}}
\newcommand{\textdiagram}[1]{\renewcommand{\usebilderscale}{\textbilderscale}\diagram{#1}\renewcommand{\usebilderscale}{\bilderscale}}
\newcommand{\diagram}[1]{\settowidth{\bilderlength}{\bilderskip\includegraphics[scale=\usebilderscale]{./#1}\bilderskip}\parbox{\bilderlength}{\bilderskip\includegraphics[scale=\usebilderscale]{./#1}\bilderskip}}

\graphicspath{{figures/},{}}

\renewcommand{\paragraph}{\subsubsection*}

\begin{document}

\bibliographystyle{KAY-hyper}

\title{{\bf Field Theories for Loop-Erased Random Walks}}
\author[ensparis]{{\bf Kay J\"org Wiese}}
\author[enslyon]{{\bf Andrei A. Fedorenko}}
\address[ensparis]
{Laboratoire de Physique de l'Ecole normale sup\'erieure, ENS, Universit\'e PSL, CNRS, Sorbonne Universit\'e, Universit\'e Paris-Diderot, Sorbonne Paris Cit\'e, 24 rue Lhomond, 75005 Paris, France.}
\address[enslyon]{Univ.\ Lyon, ENS de Lyon, Univ.\ Claude Bernard, CNRS, Laboratoire de Physique, F-69342 Lyon, France}

\begin{abstract}
Self-avoiding walks (SAWs) and   loop-erased random walks (LERWs) are two ensembles of   random
 paths with  numerous applications in mathematics, statistical physics and quantum field theory.
While   SAWs  are described by the   $n \to 0$ limit of
$\phi^4$-theory with $O(n)$-symmetry,   LERWs have no obvious field-theoretic description. We analyse
  two   candidates for a field theory of   LERWs, and   discover a   connection between
the   corresponding and a priori unrelated theories.
The first such candidate is the $O(n)$-symmetric $\phi^4$ theory at $n=-2$ whose link to   LERWs was
known in two dimensions due to conformal field theory. Here it is established  in arbitrary dimension via a perturbation
expansion in the coupling constant. The second candidate is a field theory for
charge-density waves  pinned by quenched disorder, whose relation to   LERWs had been
conjectured earlier using analogies with  Abelian sandpiles.
We explicitly show that both theories yield identical results
to 4-loop order and give both a perturbative and a non-perturbative proof of their equivalence.
This allows us to compute the fractal dimension of LERWs to order $\epsilon^5$ where $\epsilon=4-d$.
In particular, in $d=3$ our theory gives $z_{\rm LERW}(d=3)= 1.6243 \pm 0.001$,
in   excellent agreement with  the estimate  $z = 1.624 00 \pm 0.00005$ of numerical simulations.
\end{abstract}

\maketitle

\begin{figure}
\centering{}
	\includegraphics[width=0.9\columnwidth]{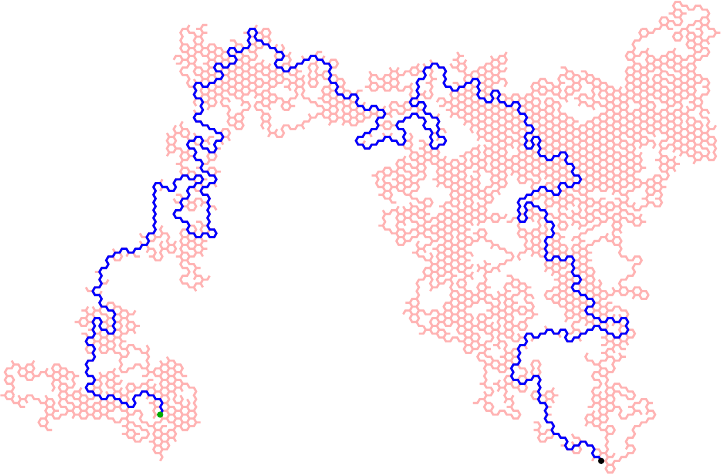}
\caption{Trace of a LERW in blue, with the erased loops in red, on a 2-dimensional Honeycomb lattice.}
\label{f:LERW}
\end{figure}

\kaysubsection{Introduction}
Random walks (RWs) which are not allowed to self-intersect
play an important role in combinatorics, statistical physics and quantum field theory. The two most
prominent models are {\em  self-avoiding walks} (SAWs) and   {\em loop-erased random walks} (LERWs).
The SAW was first introduced in polymer physics  to model long polymer chains  with self-repulsion due to excluded-volume effects.
It is defined as the uniform measure on RW paths of a given length
conditioned on having no self-intersection. Though
this model is   difficult to  analyze   rigorously,
it was discovered by de Gennes \cite{DeGennes1972} that its scaling behavior in $d$ dimensions
is given by the $O(n)$ symmetric $\phi^4$ theory in the unusual limit of $n \to 0$.
A loop-erased random walk (LERW) is defined as the trajectory
of a random walk (RW) in which any loop is erased as soon as it is formed \cite{Lawler1980}. An example is shown on figure \ref{f:LERW}, where the underlying RW is drawn in red, and the  LERW remaining after erasure in blue.
Similar to a self-avoiding walk it has a scaling limit
in all dimensions, e.g.~the end-to-end distance $R$ scales with the intrinsic length $\ell$
as $R \sim  \ell^{1/z} $, where $z$ is the fractal dimension~\cite{Kozma2007}.
It is   crucial to note that while both LERWs and SAWs are non-self-intersecting, their  fractal dimensions do not agree since they have a different statistics on the same set of allowed trajectories.
LERWs appear in many combinatorial problems, e.g.\ the shortest path on a uniform spanning tree is a LERW. While LERWs are non-Markovian RWs, their traces are equivalent to those of the {\em Laplacian Random Walk} \cite{LyklemaEvertszPietronero1986,Lawler2006}, which is Markovian, if one considers the whole trace as state variable. It is constructed on the lattice by solving the Laplace equation $\nabla^2 \Phi(x)=0$ with boundary conditions $\Phi(x)=0$ on the already constructed curve, while $\Phi(x)=1$ at the destination of the walk, either a chosen point, or infinity. The walk then advances from its tip $x$ to a neighbouring point $y$, with probability proportional to $\Phi(y)$. In a variant of this model    growth is allowed not only from the tip, but from any point on the already constructed object. This is known as the {\em dielectric breakdown model}
\cite{NiemeyerPietroneroWiesmann1984}, the simplest model for lightning. The same construction  pertains to diffusion-limited aggregation \cite{WittenSander1981}.
In contrast to SAWs,
 LERWs have no obvious field-theoretic description.
In three dimensions  LERWs have been
studied   numerically~\cite{GuttmannBursill1990,AgrawalDhar2001,Grassberger2009,Wilson2010}, while
in two dimensions  they are described by  SLE with $\kappa=2$   \cite{Schramm2000,LawlerSchrammWerner2004},   predicting a fractal dimension $z_{\rm LERW}(d=2)=\frac54$. 
Coulomb-gas techniques link this to the  2d $O(n)$-model at $n=-2$ \cite{Nienhuis1982,Duplantier1992}.
Below, we give perturbative arguments that this construction can  be done  in any dimension $d$ via the  $O(n)$-symmetric $\phi^4$ theory at $n=-2$.

Coming from a different direction,
it was conjectured in \cite{FedorenkoLeDoussalWiese2008a} that the field theory of the depinning transition
of charge-density waves (CDWs) pinned by disorder    is a field theory for LERWs.
This equivalence is   based on the conjecture of Narayan and Middleton \cite{NarayanMiddleton1994}
that   pinned CDWs  can be mapped onto the Abelian sandpile model. The connection of Abelian sandpiles with
uniform spanning trees, and thus with LERWs, was earlier established by Majumdar \cite{Majumdar1992,Dhar2006}.
Despite the lack of a proof of this equivalence, the corresponding
2-loop predictions agree with   rigorous bounds \cite{FedorenkoLeDoussalWiese2008a} and have been tested against
numerical simulations at the upper critical dimension $d=4$ in \cite{Grassberger2009}
where they correctly reproduce the leading  and subleading logarithmic corrections of \cite{FedorenkoLeDoussalWiese2008a}.
The depinning transition of   CDWs is described by the functional RG (FRG) fixed point
for random periodic systems,
proposed by Narayan and Fisher \cite{DSFisher1985,NarayanDSFisher1992b,NarayanDSFisher1992a}, and
developed in
\cite{LeschhornNattermannStepanowTang1997,NattermannStepanowTangLeschhorn1992,LeDoussalWieseChauve2003,LeDoussalWieseChauve2002,ChauveLeDoussalWiese2000a,WieseHusemannLeDoussal2018,HusemannWiese2017}.

\begin{figure}
\centering{}
\begin{center}
\fboxsep0mm
{\setlength{\unitlength}{1cm}
\begin{picture}(8.6,3.5)
\definecolor{kyellow}{rgb}{0.95,1,0.6}
\put(0,0){{\colorbox{kyellow}{\rule{86mm}{0mm}\rule{0mm}{35mm}}}}
\put(0.5,0.25){{\colorbox{red!30}{\rule{56mm}{0mm}\rule{0mm}{30mm}}}}
\put(1,1.9){{\colorbox{cyan!20}{\rule{46mm}{0mm}\rule{0mm}{10mm}}}}
\put(1.6,1.4){$\phi^{4}$-theory at $n=-2$}
\put(1.3,0.8){1 boson and 2 fermions}
\put(2.5,0.45){(complex)}
\put(2.2,2.5){one family of}
\put(1.8,2.15){complex fermions}
\put(5.9,1.75){\begin{minipage}{3cm}
\begin{center}
CDWs\\
at\\
depinning
\end{center}
\end{minipage}}
\end{picture}}
\end{center}
\caption{The different field theories for LERWs.}
\label{f:sector-illustration}
\end{figure}
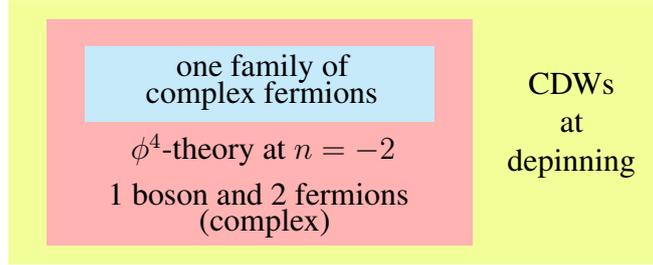

If true, both $\phi^4$-theory at $n=-2$ and the FRG for CDWs must agree, at least  for observables  related   to LERWs.
 We   show below that both the $\beta$-function and the fractal dimension of LERWs coincide. This is done   using ({\it i}) graph-theoretical arguments valid at all orders in perturbation theory,  ({\it ii})   non-perturbative supersymmetry techniques,  and  ({\it iii})    an explicit 4-loop calculation.
 This does not mean that the theories are identical:
 for example, at   depinning   CDWs exhibit  avalanches which are seemingly absent in the   $\phi^4$ model. Our statement   is that in the {\em sector}   in which we can compare the two theories, they agree. This is illustrated on figure \ref{f:sector-illustration}. It does  not exclude that one of the theories can handle observables the other cannot. A classical example for this behavior is the solution of the 2d Ising model via the conformal bootstrap, as proposed by Belavin, Polyakov and Zamolodchikov 
 (BPZ) \cite{BelavinPolyakovZamolodchikov1984}. Here a theory with three operators, the energy $\epsilon$, the spin $\sigma$, and the identity $1\!\!1$,   closes under OPE. However,   other observables   can be constructed from the Ising model on a lattice. Examples in case are domain walls, at criticality   described by SLE. On the other hand, SLE does not (at least obviously) describe the operators of the original BPZ construction. Thus, SLE and BPZ describe different {\em sectors}  of the same theory.

\kaysubsection{Mapping of {LERWs} onto  $O(n)$ $\phi^{4}$-theory at $n=-2$}
We now map LERWs onto the
 $n$-component $\phi^{4}$ theory at $n=-2$. The latter is defined by  the action
\be\label{theory1}
{\cal S}[\vec \phi] := \int_x \half [\nabla \vec \phi(x)]^2 + \frac{m^2}{2}  \vec \phi(x)^2 + \frac{g}{  8}  [\vec \phi(x)^2]^2\ .
\ee
One   checks that, for $n=-2$, the full 2-point correlation function is given in Fourier space by the free-theory result
\be\label{cor-func}
\left< \phi_i(k) \phi_j(k)\right> = \left< \phi_i(k) \phi_j(k)\right>_{0} = \frac{\delta_{{ij}}}{k^{2}+m^{2}}\ ,
\ee
independent of $g$. This fact is well known perturbatively
\cite{Zinn-JustinBook2,KleinertSchulte-FrohlindeBook,KleinertNeuSchulte-FrohlindeLarin1991,KompanietsPanzer2017}. A non-perturbative derivation is given below, by mapping onto  complex fermions.
 Eq.~(\ref{cor-func}) is the Laplace transform of the $k$-dependent Green function for a random walk (RW),
\be\label{phi-prop}
\left< \phi_i(k) \phi_j(k)\right> = \int_{0}^{\infty}\rmd t\, \rme^{-m^{2}t } \times \rme^{-k^{2 } t }\ .
\ee
Here $t\ge \ell$ is the time of the RW used to construct the  LERW of length $\ell$,  which scales as $\ell\sim t^{z/2}\sim m^{-z}$, and $z$ is the fractal dimension of the LERW.  Let us   convene that we draw the trajectory of a random walk in blue, and when it hits itself we do not erase the emerging loop, but color it in red; see  figure~\ref{f:LERW}  for an example.
We claim that we can reconstruct these {\em colored} trajectories  from $\phi^{4}$ theory.
To this aim, we first reformulate the theory \eq{theory1} in terms of $N=n/2$ complex bosons  $\Phi$ and $\Phi^{*}$, with  $\Phi_{i}(x):=\frac1{\sqrt {2}}[\phi_{2i-1}(x)+ i \phi_{2i}(x)]$, $i=1,...,N=n/2$. Its action reads
\be\label{theory2}
{\cal S}[\vec \Phi] := \!\! \int_x \!  \nabla \vec \Phi^*(x)  {\nabla  \vec\Phi}(x)  +  {m^2}  \vec \Phi^{*}(x)  \vec \Phi(x) + \frac{g}{  2}  [\vec \Phi^{*}(x) \vec \Phi(x)]^2.
\ee
This reformulation gives propagators a direction, making it easier to interpret them as RWs.
Consider a specific path with  $s$ intersections in the path-integral representation. This conditioning introduces a small-length cutoff $a$ 
 \cite{DesCloizeauxJannink,DesCloizeaux1981}: One may think of putting a grid of box-size $a$, and introduce for each box a  $\Theta$-function, which is one if the path passes through twice, and zero otherwise. It is crucial that for the self-intersections encoded in the quartic term we use the same prescription,   that it squares to one, and   is symmetric under exchange of the momenta.
 The contributions  to paths  with $s=1$ self-intersections are
{\setlength{\unitlength}{1cm}
\bea
&&
{\parbox{2.25\unitlength}{\begin{picture}(2.15,1)\put(0.15,0){\ffig{2\unitlength}{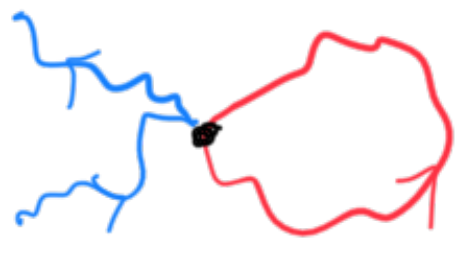}}
\put(0.,0){$\scriptstyle x$}
\put(1.0,0.7){$\scriptstyle y$}
\put(0.45,0){\scriptsize 1}
\put(1.9,0.5){\scriptsize 2}
\put(0.45,0.925){\scriptsize 3}
\put(0.03,0.95){$\scriptstyle z$}
\end{picture}}}
\ \ \ \longrightarrow \ \ \
{\parbox{1.1cm}{{\begin{tikzpicture}
\coordinate (v1) at  (0,1.25) ; \coordinate (v2) at  (0,-.25) ;  \node (x) at  (0,0)    {$\!\!\!\parbox{0mm}{$\raisebox{-3mm}[0mm][0mm]{$\scriptstyle x$}$}$};
\coordinate (x1) at  (0.5,0);\coordinate (y) at  (1.5,0.5); \coordinate (y1) at  (0.5,1) ;\node (z) at  (0,1)    {$\!\!\!\parbox{0mm}{$\raisebox{1mm}[0mm][0mm]{$\scriptstyle z$}$}$};
\fill (x) circle (1.5pt);
\fill (z) circle (1.5pt);
\draw [blue] (x) -- (x1);
\draw [blue] (y1) -- (z);
\draw [blue,directed](0.5,0) arc (-90:90:0.5);
\end{tikzpicture}}}}~
-g
{\parbox{1.6cm}{{\begin{tikzpicture}
\node (v1) at  (0,1.25){} ;
\node (v2) at  (0,-.25){} ;
\node (x) at  (0,0)    {$\!\!\!\parbox{0mm}{$\raisebox{-3mm}[0mm][0mm]{$\scriptstyle x$}$}$};
\coordinate (x1) at  (1,0) ;\coordinate (y) at  (1.5,0.5); \coordinate (y1) at  (1,1);\node (z) at  (0,1)    {$\!\!\!\parbox{0mm}{$\raisebox{1mm}[0mm][0mm]{$\scriptstyle z$}$}$};
\fill (x) circle (1.5pt);
\fill (z) circle (1.5pt);
\fill (x1) circle (1.5pt);
\fill (y1) circle (1.5pt);
\draw [blue,directed] (x) -- (x1);
\draw [blue,directed] (y1) -- (z);
\draw [blue,directed](1,0) arc (-90:90:0.5);
\draw [dashed] (x1) -- (y1);
\end{tikzpicture}}}}~
-g N
{\parbox{2.6cm}{{\begin{tikzpicture}
\node (v1) at  (0,1.25){} ;
\node (v2) at  (0,-.25){} ;
\node (x) at  (0,0)    {$\!\!\!\parbox{0mm}{$\raisebox{-3mm}[0mm][0mm]{$\scriptstyle x$}$}$};
\coordinate (x1) at  (0.5,0) ;\coordinate (y) at  (2.5,0.5);\coordinate (y1) at  (0.5,1);\coordinate (y2) at  (1.5,1) ; \node (z) at  (0,1)    {$\!\!\!\parbox{0mm}{$\raisebox{1mm}[0mm][0mm]{$\scriptstyle z$}$}$};
\coordinate (h1) at  (1,0.5) ;
\coordinate (h2) at  (1.5,0.5) ;
\fill (x) circle (1.5pt);
\fill (z) circle (1.5pt);
\fill (h1) circle (1.5pt);
\fill (h2) circle (1.5pt);
\draw [blue,directed] (x) -- (x1);
\draw [blue,directed] (y1) -- (z);
\draw [blue](0.5,0) arc (-90:90:0.5);
\draw [red,directed](1.5,0.5) arc (-180:180:0.5);
\draw [dashed] (h1) -- (h2);
\end{tikzpicture}}}}\ \ .  \label{eq:LERW-diag}
\eea}%
The drawing on the l.h.s.\ of equation~(\ref{eq:LERW-diag})
is a LERW  starting at $x$, ending in $z$, and passing through the  segments numbered 1 to 3. Due to the crossing at $y$, the loop labeled 2 is erased; we draw it in red. The r.h.s of equation~(\ref{eq:LERW-diag})  gives all   diagrams   of $\phi^4$ theory  up to order $g^{s}$.
The first term is the free-theory result, proportional to $g^{0}$. The second term $\sim g$ cancels the first term, if one puts $g\to 1$. Here it is crucial to have the same   regularization for the interaction as for the  conditioning. The third term is proportional to $N$, due to the loop, indicated in red. Setting  $N\to -1$ compensates for the
subtracted second term. Thus setting $g\to 1$ and $N\to -1$,   the probability to go from $x$ to $z$ remains  unchanged as compared to the free theory. This is a necessary condition to be    satisfied. Since the first two terms cancel, what remains is the last diagram,  corresponding   to the drawing for the trajectory of the LERW we started with.

Let us consider how this continues for $s=2$ intersections. Once a first loop has been formed, there are two possibilities: The walk can either hit a blue or a red part of its own trace. Let us first assume it hits a blue part. Then a second loop is formed, and the mapping at $g=1$, and $N=-1$ reads
\be
\!\!\!\!{\setlength{\unitlength}{1cm}\parbox{3.15\unitlength}{\begin{picture}(3.25,1)\put(0.1,0){\ffig{3\unitlength}{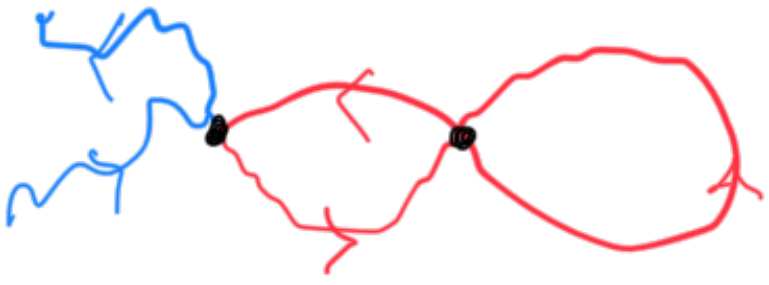}}
\put(0.,0.05){$\scriptstyle x$}
\put(0.4,0.2){\scriptsize 1}
\put(1.45,0.3){\scriptsize 2}
\put(2.99,0.6){\scriptsize 3}
\put(1.3,0.83){\scriptsize 4}
\put(0.45,1.05){\scriptsize 5}
\put(1.81,0.35){$\scriptstyle y$}
\put(0.05,0.95){$\scriptstyle z$}
\end{picture}}} \longrightarrow {\parbox{3.5cm}{{\begin{tikzpicture}
\node (v1) at  (0,1.25){} ;
\node (v2) at  (0,-.25){} ;
\coordinate (x) at  (0,0)  ;\coordinate (x1) at  (0.5,0) ;\coordinate (y) at  (2.5,0.5) ;\coordinate (k1) at  (3,0.5) ;\coordinate (k2) at  (4,0.5) ;\coordinate (y1) at  (0.5,1);\coordinate (y2) at  (1.5,1) ; \coordinate (z) at  (0,1)  ; \coordinate (h1) at  (1,0.5) ;
\coordinate (h2) at  (1.5,0.5) ;
\fill (x) circle (1.5pt);
\fill (z) circle (1.5pt);
\fill (y) circle (1.5pt);
\fill (k1) circle (1.5pt);
\fill (h1) circle (1.5pt);
\fill (h2) circle (1.5pt);
\draw [blue,directed] (x) -- (x1);
\draw [blue,directed] (y1) -- (z);
\draw [blue](0.5,0) arc (-90:90:0.5);
\draw [red,directed](2.5,0.5) arc (0:180:0.5);
\draw [red,directed](1.5,0.5) arc (180:360:0.5);
\draw [red,directed](3,0.5) arc (-180:180:0.5);
\draw [dashed] (h1) -- (h2);
\draw [dashed] (k1) -- (y);
\end{tikzpicture}}}}\hspace{0.5cm}
\ee
This is a result of multiple cancelations, which we can analyze vertex by vertex.
The tricky part is what happens at point $y$: We can either not use any interaction, use the interaction following the lines of the original walk, or reconnect the lines of the walk to form a loop. Here we   used cancelation of the first two terms, retaining   the last one, which resulted in the given drawing. (Note that the drawn red trace has the statistics of Brownian motion, as the two possible interaction terms mutually cancel.)

The other possibility is to hit a red part of the trace, say at point $y$
\be
{\setlength{\unitlength}{1cm}\parbox{3.25\unitlength}{\begin{picture}(3.25,1.2)\put(0.25,0){\ffig{3\unitlength}{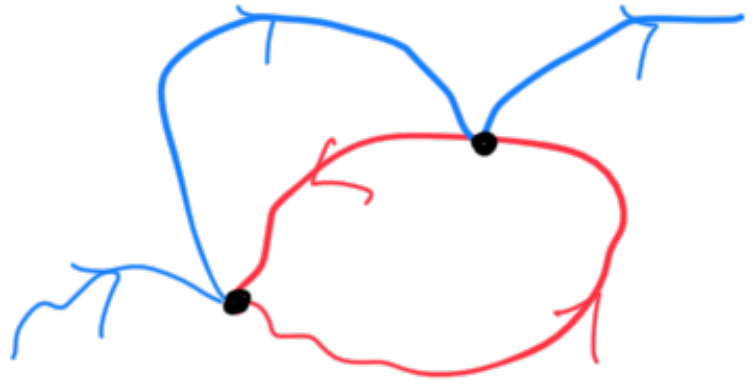}}
\put(0.1,0){$\scriptstyle x$}
\put(0.45,0){\scriptsize 1}
\put(1.9,0.2){\scriptsize 2}
\put(1.8,0.7){\scriptsize 3}
\put(0.65,1){\scriptsize 4}
\put(2.5,1.45){\scriptsize 5}
\put(2.09,.78){$\scriptstyle y$}
\put(3.25,1.4){$\scriptstyle z$}
\end{picture}}}\longrightarrow {\parbox{2.6cm}{{\begin{tikzpicture}
\node (v1) at  (0,1.25){} ;
\node (v2) at  (0,-.25){} ;
\coordinate (x) at  (0,0)  ;\coordinate (x1) at  (0.5,0) ;\coordinate (k1) at  (1,1) ;
\coordinate (k2) at  (1.5,1) ;
\coordinate (y) at  (2.5,0.5) ;\coordinate (y1) at  (0.5,1.5);\coordinate (y2) at  (1.5,1.5) ; \coordinate (z) at  (0,1.5)  ; \coordinate (h1) at  (1,0.5) ;
\coordinate (h2) at  (1.5,0.5) ;
\fill (x) circle (1.5pt);
\fill (k1) circle (1.5pt);
\fill (k2) circle (1.5pt);
\fill (z) circle (1.5pt);
\fill (h1) circle (1.5pt);
\fill (h2) circle (1.5pt);
\draw [blue,directed] (x) -- (x1);
\draw [blue,directed] (y1) -- (z);
\draw [blue](0.5,0) arc (-90:0:0.5);
\draw [blue](1.,1.0) arc (0:90:0.5);
\draw [red](2.5,1) arc (0:180:0.5);
\draw [red](1.5,0.5) arc (180:360:0.5);
\draw [dashed] (k2) -- (k1);
\draw [dashed] (h1) -- (h2);
\draw [blue,directed] (h1) -- (k1);
\draw [red,directed] (k2) -- (h2);
\draw [red,directed] (2.5,0.5)--  (2.5,1);
\end{tikzpicture}}}}~~.
\ee
Here nothing should happen, since the walk does not see the erased part of its trace. The appropriate cancelation   is between ``no interaction'' and ``reconnecting'', since the latter would  result in the erased loop appearing again in blue in perturbation theory. Thus also in this case we map onto the appropriate diagram of $\phi^4$-theory.
Continuing these arguments inductively allows us to prove this   for any number of intersections $s$.

We have thus established a one-to-one correspondence between traces of LERWs and diagrams in perturbation theory.
We still need an observable which is $1$ when inserted into a blue part of the trace, and $0$ within a red part.
This can be achieved by  modifying the probability to diffuse from $x$ to $z$, given by the expectation of $\Phi_{1}^{*}(x)\Phi_{1}(z)  $, to
\be\label{calO}
{\cal O}(x,y,z) := \Phi_{1}^{*}(x)\Phi_{1}(z) \left[ \Phi^{*} _{1}(y) \Phi _{1}(y) - \Phi^{*} _{2}(y) \Phi  _{2}(y)\right] \ .
\ee
As shown in appendix \ref{app:calO} this observable can be simplified to
\be\label{calObis}
{\cal O}(x,y,z) = \Phi_{1}^{*}(x) \Phi _{1}(y)   \Phi^{*} _{2}(y)  \Phi_{2}(z)\ .
\ee
The fractal dimension $z$ of a LERW is extracted from the length of the walk after erasure (blue part)
\be\label{10}
\frac{\left< \int_{{y,z}} {\cal O}(x,y,z) \right>}{\left< \int_{{z}} \Phi_1^{*}(x)\Phi_1(z) \right>} \equiv
m^2  {\left< \int_{{y,z}} {\cal O}(x,y,z) \right>} \sim m^{-z}\ .
\ee
We recall that the fractal dimension of SAWs is given by $z_{\mathrm{SAW}} =1/\nu$
where the critical exponent $\nu$ is taken in the limit of $n \to 0$.
Remarkably, we can deform the measure on the RWs in a continuous way by tuning the parameter $n$. This       interpolates between SAWs and LERWs, and more generally gives critical curves in the $n$-component vector model. Indeed, the operator in square brackets of (\ref{calO}) is related to the  so-called {\em crossover exponent}~\cite{Amit,Kirkham1981,ShimadaHikami2016}. It is defined as (see e.g.~\cite{Kirkham1981}, Eq.~(8))
\be
\phi_{c} = \frac{2+\gamma_{\phi\phi}-\eta }{2+\gamma_{\phi^{2} } -\eta }\equiv \left( {2+\gamma_{\phi\phi}-\eta }\right) \nu \ .
\ee
Here $\gamma_{\phi\phi}$ is the RG function for the operator $\Phi_{1}\Phi_{2}^{*}$, and $\eta$ is the anomalous exponent for the  dimension of the fields.
For the LERWs considered here, $\eta=0$ and $\nu=1/2$.
More generally \cite{Amit,Kirkham1981,KiskisNarayananVranas1993},
\be
z = d_{\rm f} = {2+\gamma_{\phi\phi}-\eta }
\ee
is the fractal dimension of a curve in the critical $O(n)$ symmetric $\phi^4$ theory.
Let us discuss what this means for LERWs and SAWs. First,
for $n=-2$, the 2-point function between two points $x$ and $z$ in the $\phi^{4}$ theory is the sum over all LERWs from $x$ to $z$, weighted by the chemical potential $m^2$ conjugate to its construction time $t$. It equals  the free propagator from $x$ to $z$, since  coloring  loops in red or erasing them does not change the propagator.
%
On the other hand, if we apply our picture to SAWs,     red loops  carry a factor of $N=0$, and only configurations without self-intersections survive. In this case the ratio \eq{10} equals 1, and the fractal dimension can be inferred from the 2-point function alone, i.e. $\phi_c(n=0) = 1$  and
$d_{\rm f} = 1/ \nu (n=0) =  z_{\mathrm{SAW}}$. Thus, the 2-point function of   $\phi^4$ theory with $n=0$ equals the sum over all SAWs connecting the two points.

Let us now turn our attention to $d=2$:
There it is known that   SLE${}_{\kappa}$ for $0 \le \kappa \le 4$ is a simple  curve.
  RWs generated by the 2-dimensional $O(n)$-symmetric $\phi^4$ theory with $n=-2\cos(4\pi/\kappa)$   are non-self-intersecting  for $-2\le n \le 2$, corresponding to $2\le \kappa\le 4$.
  For $\kappa>4$ the curves are self-intersecting. For $|n|>2$ the field theory becomes massive, and the connection breaks down. As we now know that  $n=-2$ and $n=0$ correspond to LERWs and SAWs in arbitrary dimension $d$, we conjecture that there is     a family of non-selfintersecting random walks  in arbitrary dimension $d$, parameterized by $n$, such that their measure can be continuously deformed from that of LERWs at $n=-2$ to that of SAWs at $n=0$, and beyond to $n>0$. Interestingly, the barrier of $|n|=2$  in two dimensions is seemingly absent in higher dimensions. This may be understood from the fact that the return probability of a random walk to the origin decreases with increasing dimension.


\kaysubsection{$\phi^{4}$-theory at $N=-1$, and fermions}
Up to now, we worked with $N=-1$ families of complex bosons. We   show below that instead one can consider the limit of $N_{\rm f}\to 1$ in a theory with $N_{{\rm f}}$   complex fermions, or more generally with $N_{\rm b}$ bosons and $N_{\rm f}$ fermions,  where $N=N_{\rm b}-N_{{\rm f}}$. Among others, this   provides a non-perturbative proof that
the propagator at $N=-1$ is independent of $g$.

The correspondence is based on the observation that   $N_{\rm b}$-component bosons $\vec \Phi$ carry a factor of $N_{\rm b}$ per loop, while $N_{\rm f}$-component fermions $\vec \Psi$ yield a factor of $-N_{{\rm f}}$. On a more formal level,
this can be inferred from the path integrals for   both theories. Setting
$
\mathbb{H}_{0}:= -\nabla^{2}+  \tau(x)
$
then
\bea
\!\!\!{\cal Z}_{0}^{\rm b}=
\int{\cal D}[\vec\Phi^{*}] {\cal D}[\vec\Phi] \,\rme^{-\int_{x}\vec\Phi^{*}(x)\mathbb H_{0} \vec\Phi(x)} &=& \rme^{-N_{\rm b} \tr \ln( \mathbb H_{0})} ,~~~~\\
\!\!\!{\cal Z}_{0}^{\rm f}=
\int{\cal D}[\vec\Psi^{*}]{\cal D}[\vec\Psi] \,\rme^{-\int_{x}\vec\Psi^{*}(x)\mathbb H_{0}\vec\Psi(x)} &=&  \rme^{N_{{\rm f}} \tr \ln( \mathbb H_{0})} .
\eea
The bosonic  correlation function is given by
\bea
&&\!\!\!\left< \Phi_{i}(x)\Phi_{j}^{*}(y)\right> = \delta _{{ij}}\left( \mathbb{H}_{0}^{-1}\right) _{x,y}
= \frac1{{\cal Z}_{0}^{\rm b}}\!\int{\cal D}[\vec\Phi^{*}]{\cal D}[\vec\Phi]\, \Phi_{i}(x)\Phi_{j}^{*}(y)\,\rme^{-\int_{x}\vec\Phi^{*}(x)\mathbb H_{0} \vec\Phi(x)} \ . 
\eea
For fermions  an equivalent expression holds, and  $
\left< \Psi_{i}(x) \Psi^{*}_{j}(y)\right> = \left<\Phi_{i}(x) \Phi^{*}_{j}(y)\right>
$.
Setting $\tau (x)= m^{2} + i \rho(x)$, a Hubbard Stratonovich transformation  allows us to decouple the quartic interaction   in a theory of $N_{\rm b}$ bosons  $\Phi_i$ and $N_{\rm f}$ fermions $\Psi_i$,
\bea\label{Phi-Psi-interaction}
\rme^{- \frac{g}{2}\int_x  [\vec \Phi^{*}(x) \vec \Phi(x) + \vec \Psi^{*}(x) \vec \Psi(x)]^2} 
= \int{\cal D} [\rho] \,\rme^{-\frac1{2g}\int_{x } \rho(x)^{2} - i \rho(x)[ \vec\Phi^{*}(x)\vec\Phi(x)+\vec\Psi^{*}(x)\vec\Psi(x)]}\ . 
\eea
As a consequence, a system of $N_{\rm b}$ bosons and $N_{\rm f}$ fermions with the interaction \eq{Phi-Psi-interaction}
has partition function
\bea\label{15}
{\cal Z}^{\rm b+f} &=& \int{\cal D}[\vec\Phi^{*}] {\cal D}[\vec\Phi] {\cal D}[\vec\Psi^{*}] {\cal D}[\vec\Psi]
~~~ \rme^{-\int_x \vec \Phi^*(x)(-\nabla^2+m^2) \vec \Phi(x)+ \vec \Psi^*(x)(-\nabla^2+m^2)\vec \Psi(x)  }\nn\\
&& ~~~ \times{\rme^{- \frac{g}{2}\int_x  [\vec \Phi^{*}(x) \vec \Phi(x) + \vec \Psi^{*}(x) \vec \Psi(x)]^2} }
=\int{\cal D} [\rho] \,\rme^{ (N_{\rm f}-N_{\rm b})\, \tr \ln( \mathbb H_{0})  -\frac1{2g}\int_{x } \rho(x)^{2}   }\ .
\eea
More generally,  the correlations $\left< \Phi_{1}(y)\Phi^{*}_1(x)\right>$ in complex $N$-component $\Phi^{4}$ theory  can be calculated from a  theory with $N_{\rm b}$ bosons and $N_{\rm f}$ fermions, where $N=N_{\rm b}-N_{{\rm f}}$. For   $N_{\rm b}=0$, $N_{\rm f}=1$ one gets $N=-1$, and the interaction  is $[ \vec \Psi^{*}(x) \vec  \Psi(x)]^2$. It vanishes due to the squares of Grassmann variables: This theory of {\em complex fermions},
\be\label{16}
{\cal S} ^{{N_{\rm f}=1}}_{N_{\rm b}=0}[  \Psi^*, \Psi]  =  \int_x    \nabla   \Psi^*(x)  {\nabla   \Psi}(x)  +  {m^2}    \Psi^{*}(x)   \Psi(x)\ ,
\ee
is  a free theory. It  provides a   non-perturbative proof that correlation functions of complex $\Phi^{4}$-theory at  $N=-1$ ($n=-2$ real fields) are equivalent to those of the free theory. In $d=2$ this is also  known from lattice models \cite{Nienhuis1982}. However, it does not yield the renormalization of the coupling $g$ at $N=-1$.  To obtain the latter, one has  to study $N_{\rm f}\neq 1$, and take the limit of $N_{\rm f}\to 1$ at the end.
Or one uses one family of complex bosons $N_{\rm b}=1$ and two families of complex fermions $N_{\rm f}=2$, a formulation onto which we will map   CDWs at depinning later.

Finally, care has to be taken in identifying observables in both theories: While the 2-point functions of bosonic fields are symmetric under their exchange, those of the fermionic theory are antisymmetric. As a consequence $\left<\phi_{1}\phi_{1}\right>\neq 0$, whereas  $\left<\psi_{1}\psi_{1}\right>= 0$.

{\kaysubsection{Equivalence between $\phi^{4}$-theory at $N=-1$ and CDWs at depinning}
Charge-density waves are  ground states of solids,  where the charge density is varying spatially, with a period   set to 1. Coupling these elastic objects to quenched disorder leads after averaging over disorder to the  dynamic field theory \cite{FukuyamaLee1978,LeeRice1979,DSFisher1985,NarayanDSFisher1992b,NarayanDSFisher1992a,NattermannScheidl2000}
\be\label{CDW-action}
{\cal S}^{\rm CDW} = \int_{x,t} \tilde u (x,t)  (\partial_{t}-\nabla^{2}+m^2)   u
(x,t)
-\half \int_{x,t,t'} \tilde u (x,t)\tilde u
(x,t')  \Delta \big(u (x,t)-u (x,t')\big) .
\ee
The function $\Delta(u)$ is an even function with period 1. Its renormalization can be studied using functional RG (FRG) \cite{DSFisher1985,NarayanDSFisher1992a,NarayanDSFisher1992b,LeschhornNattermannStepanowTang1997,NattermannStepanowTangLeschhorn1992,LeDoussalWieseChauve2003,LeDoussalWieseChauve2002,ChauveLeDoussalWiese2000a}.
The analysis of the FRG flow for the function $\Delta(u)$
shows that the   fixed point has the form
\be\label{18}
\Delta(u) = \Delta(0) - \frac g{ 2} u(1-u)\ .
\ee
In the absence of higher-order terms in $u$, the RG flow closes in the   space of polynomials of degree 2, and for the quadratic term one is left with the renormalization of a single coupling constant $g$.
Thus, as long as one is not interested in 2-point correlation functions, or avalanches, the fixed-point function can be replaced by
$
\Delta (u) \to \frac g{ 2} u^2\ ,
$
which generates the  simpler field theory,
\be\label{SCDWsimp}
{\cal S}^{\rm CDW}_{\rm simp} := \int_{x,t} \tilde u (x,t)  (\partial_{t}-\nabla^{2}+m^2)   u
(x,t)
-\frac g { 4} \int_{x,t,t'} \tilde u (x,t)\tilde u
(x,t')  \big[u (x,t)-u (x,t')\big]^2\ . 
\ee
Let us define a further variant which retains from $\Delta \big(u (x,t')-u (x,t)\big) $ only the term   $ u (x,t) u (x,t')$,
\be\label{SSAW}
{\cal S}^{\rm SAW} := \int_{x,t} \tilde u (x,t)   (\partial_{t}-\nabla^{2}+m^2)   u
(x,t)  
+ { \frac g2} \int_{x,t,t'}\tilde u (x,t) u(x,t) \tilde u (x,t') u (x,t') \ . 
\ee
Perturbation expansion in this theory looks exactly like the one in \Eq{theory2}, with   a different propagator to be compared with \Eq{phi-prop},
\be\label{Rdef}
R(k,t):=\left< \tilde u (k,0) u(-k,t ) \right> = \Theta(t ) \,\rme^{-t(k^2+m^2)}\ .
\ee
In this theory closed loops have weight zero, as they are acausal   in It\^o discretization. If one can integrate freely over all times, diagrams in the dynamic theory reduce to those in the complex scalar theory.  Thus the theory defined by \Eq{SSAW} can be mapped onto the  action \eq{theory2} with
$
n=N=0
$,
i.e.\ a self-avoiding walk. This is  well-known   due to de~Gennes \cite{DeGennes1972}.

We now   show that the action \eq{SCDWsimp} has the same effective coupling as the action \eq{theory2} at $N=-1$. We   first remark that   the renormalized coupling is   extracted from     diagrams with    times   on which they depend taken   far apart.  An example is given by the first diagram in \Eq{21} below.
 To show the equivalence, we start by   drawing all diagrams   present in the SAW-theory \eq{SSAW}, complementing them by the missing diagrams originating from the additional vertices of \eq{SCDWsimp} as compared to \eq{SSAW}.
These missing
diagrams, a.k.a.\ {\em children}, can be generated from the diagrams for SAWs by moving one arrow from one side of the
vertex into which it enters to the other side,
\be\label{21}
{\parbox{2cm}{{\begin{tikzpicture}
\coordinate (i1) at  (0.,0) ;
\coordinate (i2) at  (0.,1) ;
\coordinate (o1) at  (2.,0) ;
\coordinate (o2) at  (2.,1) ;
\coordinate (x1) at  (0.5,0);
\coordinate (x2) at  (1.5,0);
\coordinate (y1) at  (0.5,1);
\coordinate (y2) at  (1.5,1);
\fill (x1) circle (1.5pt);
\fill (y1) circle (1.5pt);
\fill (x2) circle (1.5pt);
\fill (y2) circle (1.5pt);
\draw [black] (i1) -- (o1);
\draw [black] (i2) -- (o2);
\draw [black,directed] (x1) -- (x2);
\draw [black,directed] (y1) -- (y2);
\draw [black,enddirected] (x2) -- (o1);
\draw [black,enddirected] (y2) -- (o2);
\draw [dashed] (x1) -- (y1);
\draw [dashed] (x2) -- (y2);
\end{tikzpicture}}}}
\longrightarrow - {\parbox{2cm}{{\begin{tikzpicture}
\coordinate (i1) at  (0.,0) ;
\coordinate (i2) at  (0.,1) ;
\coordinate (o1) at  (2.,0) ;
\coordinate (o2) at  (2.,1) ;
\coordinate (x1) at  (0.5,0);
\coordinate (x2) at  (1.5,0);
\coordinate (y1) at  (0.5,1);
\coordinate (y2) at  (1.5,1);
\fill (x1) circle (1.5pt);
\fill (y1) circle (1.5pt);
\fill (x2) circle (1.5pt);
\fill (y2) circle (1.5pt);
\draw [black] (i1) -- (x1);
\draw [black] (i2) -- (y1);
\draw [black,directed] (x1) -- (x2);
\draw [red,directed] (y1) -- (x2);
\draw [black,enddirected] (x2) -- (o1);
\draw [black,enddirected] (y2) -- (o2);
\draw [dashed] (x1) -- (y1);
\draw [dashed] (x2) -- (y2);
\end{tikzpicture}}}}\ .
\ee
We then  extract contributions to the effective coupling; this is   cleverly   done by remarking that
{({\it i})} the form of the effective interaction is proportional to the second line of \Eq{SCDWsimp}, and ({\it ii}) it is extracted by retaining only terms of the form present in \Eq{SSAW}. This implies that the second diagram   in \Eq{21} does not contribute.
Indeed it comes  with two other ones, \be
-{\parbox{2cm}{{\begin{tikzpicture}
\coordinate (i1) at  (0.,0) ;
\coordinate (i2) at  (0.,1) ;
\coordinate (o1) at  (2.,0) ;
\coordinate (o2) at  (2.,1) ;
\coordinate (x1) at  (0.5,0);
\coordinate (x2) at  (1.5,0);
\coordinate (y1) at  (0.5,1);
\coordinate (y2) at  (1.5,1);
\fill (x1) circle (1.5pt);
\fill (y1) circle (1.5pt);
\fill (x2) circle (1.5pt);
\fill (y2) circle (1.5pt);
\draw [black] (i1) -- (x1);
\draw [black] (i2) -- (y1);
\draw [black,directed] (x1) -- (x2);
\draw [black,directed] (y1) -- (x2);
\draw [black,enddirected] (x2) -- (o1);
\draw [black,enddirected] (y2) -- (o2);
\draw [dashed] (x1) -- (y1);
\draw [dashed] (x2) -- (y2);
\end{tikzpicture}}}}+\frac12
{\parbox{2cm}{{\begin{tikzpicture}
\coordinate (i1) at  (0.,0) ;
\coordinate (i2) at  (0.,1) ;
\coordinate (i2bis) at  (0.,0.7) ;
\coordinate (o1) at  (2.,0) ;
\coordinate (o2) at  (2.,1) ;
\coordinate (x1) at  (0.5,0);
\coordinate (x2) at  (1.5,0);
\coordinate (y1) at  (0.5,1);
\coordinate (y2) at  (1.5,1);
\fill (x1) circle (1.5pt);
\fill (y1) circle (1.5pt);
\fill (x2) circle (1.5pt);
\fill (y2) circle (1.5pt);
\draw [black] (i2) -- (y1);
\draw [black] (i2bis) -- (y1);
\draw [black,directed] (x1) -- (x2);
\draw [black,directed] (y1) -- (x2);
\draw [black,enddirected] (x2) -- (o1);
\draw [black,enddirected] (y2) -- (o2);
\draw [dashed] (x1) -- (y1);
\draw [dashed] (x2) -- (y2);
\end{tikzpicture}}}}
+\frac12 {\parbox{2cm}{{\begin{tikzpicture}
\coordinate (i1) at  (0.,0) ;
\coordinate (i1bis) at  (0.,0.3) ;
\coordinate (i2) at  (0.,1) ;
\coordinate (o1) at  (2.,0) ;
\coordinate (o2) at  (2.,1) ;
\coordinate (x1) at  (0.5,0);
\coordinate (x2) at  (1.5,0);
\coordinate (y1) at  (0.5,1);
\coordinate (y2) at  (1.5,1);
\fill (x1) circle (1.5pt);
\fill (y1) circle (1.5pt);
\fill (x2) circle (1.5pt);
\fill (y2) circle (1.5pt);
\draw [black] (i1) -- (x1);
\draw [black] (i1bis) -- (x1);
\draw [black,directed] (x1) -- (x2);
\draw [black,directed] (y1) -- (x2);
\draw [black,enddirected] (x2) -- (o1);
\draw [black,enddirected] (y2) -- (o2);
\draw [dashed] (x1) -- (y1);
\draw [dashed] (x2) -- (y2);
\end{tikzpicture}}}}\ .
\ee
After time-integration, the two fields at the left-most vertex     cancel, thus
the above sum   vanishes.
The next diagram
\be
{\parbox{2.2cm}{{\begin{tikzpicture}
\coordinate (i1) at  (0.,0) ;
\coordinate (i2) at  (0.,1) ;
\coordinate (o1) at  (2.,0) ;
\coordinate (o2) at  (2.,1) ;
\coordinate (x1) at  (1.5,0) ;
\coordinate (x2) at  (2.,0) ;
\coordinate (y1) at  (1.5,1);
\coordinate (y2) at  (2.,1) ;
\coordinate (m1) at  (0.5,0.5) ;
\coordinate (m2) at  (1.,0.5) ;
\fill (m1) circle (1.5pt);
\fill (m2) circle (1.5pt);
\fill (x1) circle (1.5pt);
\fill (y1) circle (1.5pt);
\draw [black] (y2) -- (y1);
\draw [black,enddirected] (x1) -- (x2);
\draw [black,enddirected](0.,0) arc (-90:90:0.5);
\draw [black,directed](1.5,1) arc (90:180:0.5);
\draw [black,directed](1,0.5) arc (180:270:0.5);
\draw [dashed] (x1) -- (y1);
\draw [dashed] (m1) -- (m2);
\end{tikzpicture}}}}
\ee
has two children,
\be
-{\parbox{2.2cm}{{\begin{tikzpicture}
\coordinate (i1) at  (0.,0) ;
\coordinate (i2) at  (0.,1) ;
\coordinate (o1) at  (2.,0) ;
\coordinate (o2) at  (2.,1) ;
\coordinate (x1) at  (1.5,0) ;
\coordinate (x2) at  (2.,0) ;
\coordinate (y1) at  (1.5,1);
\coordinate (y2) at  (2.,1) ;
\coordinate (m1) at  (0.5,0.5) ;
\coordinate (m2) at  (1.,0.5) ;
\fill (m1) circle (1.5pt);
\fill (m2) circle (1.5pt);
\fill (x1) circle (1.5pt);
\fill (y1) circle (1.5pt);
\draw [black] (y2) -- (y1);
\draw [black,enddirected] (x1) -- (x2);
\draw [black,enddirected](0.,0) arc (-90:90:0.5);
\draw [black,directed](1.5,1) arc (90:180:0.5);
\draw [red,directed](1,0.5) arc (-90:0:0.5);
\draw [dashed] (x1) -- (y1);
\draw [dashed] (m1) -- (m2);
\end{tikzpicture}}}}
-{\parbox{2.2cm}{{\begin{tikzpicture}
\coordinate (i1) at  (0.,0) ;
\coordinate (i2) at  (0.,1) ;
\coordinate (o1) at  (2.,0) ;
\coordinate (o2) at  (2.,1) ;
\coordinate (x1) at  (1.5,0) ;
\coordinate (x2) at  (2.,0) ;
\coordinate (y1) at  (1.5,1);
\coordinate (y2) at  (2.,1) ;
\coordinate (m1) at  (0.5,0.5) ;
\coordinate (m2) at  (1.,0.5) ;
\fill (m1) circle (1.5pt);
\fill (m2) circle (1.5pt);
\fill (x1) circle (1.5pt);
\fill (y1) circle (1.5pt);
\draw [black] (y2) -- (y1);
\draw [black,enddirected] (x1) -- (x2);
\draw [black,enddirected](0.,0) arc (-90:90:0.5);
\draw [red,directed] (y1) -- (m1);
\draw [black,directed](1,0.5) arc (180:270:0.5);
\draw [dashed] (x1) -- (y1);
\draw [dashed] (m1) -- (m2);
\end{tikzpicture}}}}\to 0\ .
\ee
They both vanish, since the first is an acausal loop, thus zero, and the second has only one connected component, thus does not contribute to the renormalization of $g$.
Now consider
\be
{\parbox{3.2cm}{{\begin{tikzpicture}
\coordinate (i1) at  (0.,0)  ;
\coordinate (i2) at  (0.,1) ;
\coordinate (o1) at  (2.5,0.5) ;
\coordinate (o2) at  (2,0.5);
\coordinate (x1) at  (1.5,0);
\coordinate (x2) at  (2.,0);
\coordinate (y1) at  (1.5,1);
\coordinate (y2) at  (2.,1) ;
\coordinate (m1) at  (0.5,0.5) ;
\coordinate (m2) at  (1.,0.5) ;
\fill (m1) circle (1.5pt);
\fill (m2) circle (1.5pt);
\fill (o1) circle (1.5pt);
\fill (o2) circle (1.5pt);
\draw [black,enddirected](0.,0) arc (-90:90:0.5);
\draw [black,directed](2,0.5) arc (0:180:0.5);
\draw [black,directed](1,0.5) arc (180:360:0.5);
\draw [black,enddirected](3.,1) arc (90:270:0.5);
\draw [dashed] (o1) -- (o2);
\draw [dashed] (m1) -- (m2);
\end{tikzpicture}}}}
\ee
This   diagram contains an acausal loop, thus does not contribute to the SAW theory \eq{SSAW} where it vanishes due to a factor of $N=0$. The diagram however has
 {\em children}; together they are
\bea\label{28}
{\parbox{3.2cm}{{\begin{tikzpicture}
\coordinate (i1) at  (0.,0)  ;
\coordinate (i2) at  (0.,1) ;
\coordinate (o1) at  (2.5,0.5) ;
\coordinate (o2) at  (2,0.5);
\coordinate (x1) at  (1.5,0);
\coordinate (x2) at  (2.,0);
\coordinate (y1) at  (1.5,1);
\coordinate (y2) at  (2.,1) ;
\coordinate (m1) at  (0.5,0.5) ;
\coordinate (m2) at  (1.,0.5) ;
\fill (m1) circle (1.5pt);
\fill (m2) circle (1.5pt);
\fill (o1) circle (1.5pt);
\fill (o2) circle (1.5pt);
\draw [black,enddirected](0.,0) arc (-90:90:0.5);
\draw [black,directed](2,0.5) arc (0:180:0.5);
\draw [black,directed](1,0.5) arc (180:360:0.5);
\draw [black,enddirected](3.,1) arc (90:270:0.5);
\draw [dashed] (o1) -- (o2);
\draw [dashed] (m1) -- (m2);
\end{tikzpicture}}}}
&-&{\parbox{3.2cm}{{\begin{tikzpicture}
\coordinate (i1) at  (0.,0)  ;
\coordinate (i2) at  (0.,1) ;
\coordinate (o1) at  (2.5,0.5) ;
\coordinate (o2) at  (2,0.5);
\coordinate (x1) at  (1.5,0);
\coordinate (x2) at  (2.,0);
\coordinate (y1) at  (1.5,1);
\coordinate (y2) at  (2.,1) ;
\coordinate (m1) at  (0.5,0.5) ;
\coordinate (m2) at  (1.,0.5) ;
\fill (m1) circle (1.5pt);
\fill (m2) circle (1.5pt);
\fill (o1) circle (1.5pt);
\fill (o2) circle (1.5pt);
\draw [black,enddirected](0.,0) arc (-90:90:0.5);
\draw [red,directed](2,0.5) arc (0:180:0.75);
\draw [black,directed](1,0.5) arc (180:360:0.5);
\draw [black,enddirected](3.,1) arc (90:270:0.5);
\draw [dashed] (o1) -- (o2);
\draw [dashed] (m1) -- (m2);
\end{tikzpicture}}}}
\nn\\
 \;\;-{\parbox{3.2cm}{{\begin{tikzpicture}
\coordinate (i1) at  (0.,0)  ;
\coordinate (i2) at  (0.,1) ;
\coordinate (o1) at  (2.5,0.5) ;
\coordinate (o2) at  (2,0.5);
\coordinate (x1) at  (1.5,0);
\coordinate (x2) at  (2.,0);
\coordinate (y1) at  (1.5,1);
\coordinate (y2) at  (2.,1) ;
\coordinate (m1) at  (0.5,0.5) ;
\coordinate (m2) at  (1.,0.5) ;
\fill (m1) circle (1.5pt);
\fill (m2) circle (1.5pt);
\fill (o1) circle (1.5pt);
\fill (o2) circle (1.5pt);
\draw [black,enddirected](0.,0) arc (-90:90:0.5);
\draw [black,directed](2,0.5) arc (0:180:0.5);
\draw [red,directed](1,0.5) arc (180:360:0.75);
\draw [black,enddirected](3.,1) arc (90:270:0.5);
\draw [dashed] (o1) -- (o2);
\draw [dashed] (m1) -- (m2);
\end{tikzpicture}}}}
&+&{\parbox{3.2cm}{{\begin{tikzpicture}
\coordinate (i1) at  (0.,0)  ;
\coordinate (i2) at  (0.,1) ;
\coordinate (o1) at  (2.5,0.5) ;
\coordinate (o2) at  (2,0.5);
\coordinate (x1) at  (1.5,0);
\coordinate (x2) at  (2.,0);
\coordinate (y1) at  (1.5,1);
\coordinate (y2) at  (2.,1) ;
\coordinate (m1) at  (0.5,0.5) ;
\coordinate (m2) at  (1.,0.5) ;
\fill (m1) circle (1.5pt);
\fill (m2) circle (1.5pt);
\fill (o1) circle (1.5pt);
\fill (o2) circle (1.5pt);
\draw [black,enddirected](0.,0) arc (-90:90:0.5);
\draw [red,directed](2,0.5) arc (0:180:0.75);
\draw [red,directed](1,0.5) arc (180:360:0.75);
\draw [black,enddirected](3.,1) arc (90:270:0.5);
\draw [dashed] (o1) -- (o2);
\draw [dashed] (m1) -- (m2);
\end{tikzpicture}}}}\ .\qquad
\eea
The modified lines are in red. We first remark   that none of them restricts the time-difference between the left and right-most vertices, and they all contribute to the effective coupling.
Their coefficients are $0\times 1-1-1+1 = -1$. Integrating over times, the  result is the same as in  $\phi^4$-theory at $N=-1$,  graphically represented as
\be
 {\parbox{3.2cm}{{\begin{tikzpicture}
\coordinate (i1) at  (0.,0)  ;
\coordinate (i2) at  (0.,1) ;
\coordinate (o1) at  (2.5,0.5) ;
\coordinate (o2) at  (2,0.5);
\coordinate (x1) at  (1.5,0);
\coordinate (x2) at  (2.,0);
\coordinate (y1) at  (1.5,1);
\coordinate (y2) at  (2.,1) ;
\coordinate (m1) at  (0.5,0.5) ;
\coordinate (m2) at  (1.,0.5) ;
\fill (m1) circle (1.5pt);
\fill (m2) circle (1.5pt);
\fill (o1) circle (1.5pt);
\fill (o2) circle (1.5pt);
\draw [blue,enddirected](0.,0) arc (-90:90:0.5);
\draw [red,directed](2,0.5) arc (0:180:0.5);
\draw [red,directed](1,0.5) arc (180:360:0.5);
\draw [blue,enddirected](3.,1) arc (90:270:0.5);
\draw [dashed] (o1) -- (o2);
\draw [dashed] (m1) -- (m2);
\end{tikzpicture}}}}\ . \ee
Noting the pairwise cancelations in loops of the form \eq{28}, this can be interpreted as the {\em missing} contribution of the first (acausal) diagram.
To summarize: We    showed that at   1-loop order the action \eq{SCDWsimp} has the same effective coupling as the action \eq{theory2}, diagram by diagram (after time integration). These considerations can be formalized to higher orders, and we   checked them explicitly   up to 4 loops.

The theory \eq{SCDWsimp} has a second renormalization, namely of friction, or time, which shows up in a renormalization of $
\int_{x,t} \tilde u(x,t) \dot u(x,t).
$
The standard route to study this is to write down all diagrams constructed from \eq{SCDWsimp}, in which one field $\tilde u$ and one field $u$ remain \cite{LeDoussalWieseChauve2002}. Due to the structure of the action, the latter has the form $u(x,t)-u(x,t')$ and can be expanded as $\dot u(x,t) (t-t')$. The additional time difference, when appearing  together with a response function, acts by  inserting an additional point into the latter, as can be seen from the definition \eq{Rdef}, and the relation \be
 t R(k,t) =  \int_0^t \rmd{t'} R(k,t') R(k,t-t')\ .
\ee
Following this strategy, we checked that up to 4-loop order all diagrams appearing after time-integration are equivalent to those encountered in expectations of ${\cal O} $, defined in \Eq{calO}. Graphically, this is proven by first realizing that   one can alternatively study the  renormalization of friction by considering insertions of $\int_{x,t}\tilde u(x,t) \dot u(x,t)$.
Doing this, the time derivative on $\dot u$ can be passed through a closed string of response functions to either the earliest time in the diagram, and will then act on the remaining uncontracted field $u$, or it will end on a vertex with no further $u$ field, and thus vanish. The same argument can be done by moving the time derivative to the field $\tilde u$. These operations  restrict the class of diagrams. Graphically, inserting $\int_{x,t}\tilde u(x,t) \dot u(x,t)$ corresponds again to inserting a point into diagrams correcting expectations of $\tilde u(x,t) u(x',t')$. The final step of the proof is to realize that this is equivalent to insertions of the crossover operator $\Phi _{1}(y)   \Phi^{*} _{2}(y)$ in the theory \eq{theory2}.
}

Finally note that the absence of a renormalization of $-\nabla^2+m^2$ in Eqs.~\eq{CDW-action} and \eq{SCDWsimp}
implied by the statistical tilt symmetry is equivalent to the absence of a renormalization of the theory \eq{16}.

\kaysubsection{A non-perturbative proof for the equivalance of $\phi^4$-theory at $N=-1$ and CDWs}
The 
method introduced in \cite{ParisiSourlas1979,ParisiSourlas1982}
allows one to write the disorder average of any observable ${\cal O}[u_i]$ as
\begin{align}\label{su5}
&\overline {{\cal O}[u_i]}  = \int\prod_{a=1}^r {\cal D}[{\tilde u_{a} }] {\cal D}[{
u_{a} }] {\cal D}[{\bar \psi_{a} }] {\cal D}[{\psi}_{a}]\, {{\cal O}[u_i]}
 \times \nn \\ & \hspace{2.5cm}~~~\times
\!\overline{\,\exp\!\left[-{ \int_{x}\tilde u_{a} (x)\frac{\delta{\cal H}[{u_{a} }]} {\delta {u_{a} } (x) }+\bar \psi_{a}(x)\frac{\delta^{2}
{\cal H}[{u_{a} } ] }{\delta {u_{a} } (x)\delta {u_{a} }
(y) } \psi_{a} (y) } \right]} . 
\end{align}
The integral over $\tilde u_a$ ensures that  $u_a$ is at a minimum. $\bar \psi_a$ and $\psi_a$ are fermionic degrees of freedom (Grassmann variables), which compensate for the functional determinant appearing in the integration over $u$, yielding a partition function ${\cal Z}=1$.
Averaging over disorder gives an effective action \cite{Wiese2004}
\begin{eqnarray}\label{su6a}
\label{su6b}
&&{\cal S}[\tilde u_a, u_{a},\bar \psi_{a}, \psi_{a}]
= \!
\sum_{a}\! \int_{x} \!\tilde u_{a} (x) (-\nabla^{2}{+}m^2) u_{a} (x)  
 + \bar \psi_{a} (x)
(-\nabla^{2}{+}m^2)\psi_{a} (x) \nonumber
\\
&&- \sum_{a,b} \int_x\Big[ \half \tilde u_{a} (x)\Delta \big(u_{a} (x)-u_{b}
(x)\big)\tilde u_{b} (x)
- \tilde u_{a} (x)
\Delta' \big(u_{a} (x)-u_{b} (x)\big) \bar \psi_{b} (x)\psi_{b} (x)\nonumber \\
&& \qquad~~~   -\half \bar \psi_{a} (x)\psi _{a} (x)\Delta''
\big(u_{a} (x)-u_{b} (x)\big)\bar \psi_{b} (x)\psi_{b} (x)  \Big]. 
\end{eqnarray}
The function $\Delta(u)$ is the same as in \Eq{CDW-action}.
Note that we   allow for an arbitrary number of replicas $r$. In the   work \cite{ParisiSourlas1979} the focus was on $r=1$, which does not allow to extract the second cumulant of the disorder, i.e.\ its correlations. To do the latter, one needs at least $r=2$   copies, to which we specify now.
We   introduce center-of-mass coordinates,
\begin{align}
u_1(x) &= u(x) + \frac12 \phi(x)\ , & u_2(x)&= u(x) - \frac12  \phi(x)\ ,\\
\tilde u_1(x) &= \frac12 \tilde u(x) + \tilde \phi(x)\ , & \tilde u_2(x)&= \frac12  \tilde u(x) -  \tilde \phi (x)\ .
\end{align}
The action \eq{su6b} can then be written as
\begin{align}
\label{CDW=phi4bis} {\cal S} 
=   \int_{x} &\tilde \phi(x) (-\nabla^2 +m^2)\phi(x)+ \tilde  u(x) (-\nabla^2 +m^2) u(x)
+   \sum_{a=1}^2\bar \psi_{a} (x)
(-\nabla^{2}+m^2)\psi_{a} (x)\nn\\
&+  \tilde \phi(x)^2 \Big[ \Delta\big(\phi(x)\big)-\Delta(0)\Big] -\frac14 \tilde u(x)^2\Big[ \Delta\big(\phi(x)\big)+\Delta(0)\Big]
\nn\\ &
+ \frac12 \tilde u(x) \Delta'\big(\phi(x)\big)\Big[\bar \psi_{2}(x) \psi_{2}(x)-\bar \psi_{1}(x) \psi_{1}(x)\Big] \nn\\& +  \tilde \phi(x) \Delta'\big(\phi(x)\big)\Big[\bar \psi_{2}(x) \psi_{2}(x)+\bar \psi_{1}(x) \psi_{1}(x)\Big]
\nn\\ &
+ \bar \psi_{2}(x) \psi_{2}(x) \bar \psi_{1}(x) \psi_{1}(x) \Delta''\big(\phi(x)\big)\ .
 \end{align}
As in the derivation of the action \eq{SCDWsimp}  replacing $\Delta(u)\to \frac{g}2 u^2$, the action \eq{CDW=phi4bis} takes the form
\begin{align}
\label{CDW=phi4} {\cal S} 
=   \int_{x} &\tilde \phi(x) (-\nabla^2 +m^2)\phi(x)+ \tilde  u(x) (-\nabla^2 +m^2) u(x)
+   \sum_{a=1}^2\bar \psi_{a} (x)
(-\nabla^{2}+m^2)\psi_{a} (x) \nonumber \\
&+
 \frac g 2\tilde u(x) \phi(x)\!\Big[\bar \psi_2(x)\psi_2(x) -\bar \psi_1(x)\psi_1(x) \Big]   - \frac g 8 \tilde u(x) ^2 \phi(x)^2
 \nonumber \\ &
+ \frac g 2\left[ \tilde \phi(x)\phi(x)+\bar \psi_1(x)\psi_1(x) +\bar \psi_2(x)\psi_2(x) \right]^2 \ .
 \end{align}
Note that only $\tilde u(x)$, but not the center-of-mass  position $u(x)$ appear in the interaction. While $u(x)$  may have non-trivial expectations, it does not contribute to the renormalization of $g$, and the latter can be obtained by dropping the third  line of \Eq{CDW=phi4}. What remains  is a $\phi^4$-type theory as in \Eq{15}, with one complex boson, and two complex fermions. It can equivalently be viewed as complex $\phi^4$-theory at $N=-1$, or real $\phi^4$-theory at $n=-2$.

What is yet missing is information about the exponent $z$. One   can use the operator $\cal O$ defined in Eqs.~\eq{calO} or \eq{calObis}, replacing $\Phi_i$ by $\psi_i$, and $\Phi_i^*$ by $\bar \psi_i$. Another possibility is to introduce time, adding a time argument to all fields, and replacing   $-\nabla^2 +m^2$ by $\partial_t -\nabla^2 +m^2$.
The interaction part, i.e.\ the last line of \Eq{CDW=phi4}, then  becomes bilocal in time, i.e.\ the time integral appears inside the square bracket.
The tricky part is to ensure that time-integrated vacuum bubbles retain their static expectations. This can be done by specifying an initial condition, once again adding the action \eq{CDW=phi4} where all fields are evaluated at some initial time $t_0$. This implies that
\be
R(k;t',t)=\left< \phi(-k,t') \tilde \phi(k,t) \right> = \left< \psi_i(-k,t') \bar \psi_i(k,t) \right>
= \Theta(t'-t) \rme^{-(k^2+m^2)(t'-t)}+ \frac{\delta_{t,t_0}\delta_{t',t_0}}{k^2+m^2}\ . 
\ee
The $\delta$-functions are to be understood  s.t.
\be
\int_t R(k_1,t,t) ... R(k_n,t,t) = \frac1{(k_1^2+m^2)... (k_n^2+m^2)}\ .
\ee
We   explicitly checked to 3-loop order that   terms proportional to $\partial_t $ receive the same renormalization as at depinning. Furthermore we can analyze the renormalization of $\tilde \phi(x,t)\partial_t \phi(x,t)$ as an insertion. Contributing  diagrams carry two external fields, one  $\tilde \phi$, and one  $\phi$.  Passing the time derivative   $\partial_t \phi(x,t)$ of the insertion to this external field, what remains is the insertion of a single point in the line connecting the external fields $\tilde \phi$ and $\phi$, but no contribution from insertions into loops. After integration over   times, this is equivalent to the insertion of $\cal O$ defined  above in Eqs.~\eq{calO} or~\eq{calObis}.

\kaysubsection{Fractal dimension of LERWs at 5-loop order}
We   generated all diagrams entering into ${\cal O}(x,y,z)$ at 5-loop order, and into the renormalization of the coupling constant at $4$-loop order, supplemented by the diagrams of \cite{KompanietsPanzer2017} at 5-loop order. At 3-loop order we obtain the fractal dimension $z$ of LERWs  using the massive diagrams of Ref.~\cite{WieseHusemannLeDoussal2018}. To 4- and 5-loop order, we use diagrams in a massless minimal subtraction scheme obtained in \cite{KleinertNeuSchulte-FrohlindeLarin1991,KompanietsPanzer2017}. The result reproduces at 4-loop order the one given for the crossover exponent  in Ref.~\cite{Kirkham1981}, setting there $n=-2$.
This yields for the fractal dimension $  z$ of LERWs in dimension $d=4-\epsilon$, equivalent to the dynamical exponent of CDWs at depinning,
\begin{eqnarray} \nn
\!\!\!z &=&  2-\frac{\varepsilon }{3}- \frac{\varepsilon^2}{9}
+ \bigg[\frac{2 \zeta (3)}{9}-\frac{1}{18}\bigg]\varepsilon^3
- \bigg[\frac{70 \zeta (5)}{81} -\frac{\zeta(4)}{6} -\frac{17 \zeta (3) }{162}
   +\frac{7}{324} \bigg]\varepsilon^4 \nn\\
   &&     + \bigg[
   \frac{121 \zeta (3)}{972} -\frac{8 \zeta (3)^2}{81}+\frac{17
   \zeta (4)}{216}-\frac{103 \zeta (5)}{243}-\frac{175 \zeta
   (6)}{162}
   +\frac{833 \zeta (7)}{216}-\frac{17}{1944} \bigg]\varepsilon^5 + \,{\cal O}(\varepsilon^6) \nn\\
   &= & 2-\frac\epsilon 3  -\frac {\epsilon ^2}9 +0.211568
   \epsilon ^3-0.611186 \epsilon ^4+2.43354
   \epsilon ^5+ \,{\cal O}(\varepsilon^6)\ . \label{eq:On-24} \ \ \ \ \
\end{eqnarray}
Using Borel resummation of (\ref{eq:On-24}), where to improve the precision we have included the 6-loop term  yields \cite{KompanientsWiese2019}
\begin{eqnarray}
&& z(d=2) = 1.244\pm 0.01, \\
&& z(d=3)= 1.6243 \pm 0.001.
\end{eqnarray}
This can be compared to the exact value $z(d=2)=5/4$  \cite{Schramm2000,LawlerSchrammWerner2004}
and the most precise numerical simulations to date by David Wilson \cite{Wilson2010},
\be
z(d=3) = 1.624 00 \pm 0.000 05\ .
\ee

\medskip

\kaysubsection{Summary and Perspectives}
We   presented evidence that both $\phi^4$ theory with  $O(n)$-symmetry at $n=-2$, as well as the field theory for CDWs at depinning describe loop-erased random walks. We   sketched a proof of this equivalence, based on a diagrammatic expansion, and gave an algebraic proof for the latter. All claims were explicitly checked   to 4-loop order.
This equivalence gives a strong support for the Narayan-Middleton conjecture \cite{NarayanMiddleton1994} that CDWs pinned by disorder can be mapped onto the Abelian sandpile model, and  thus  for  the
conjecture of \cite{FedorenkoLeDoussalWiese2008a}.
Remarkably, while CDWs at depinning map onto Abelian sandpiles, disordered elastic interfaces at depinning   map onto Manna sandpiles \cite{LeDoussalWiese2014a,Wiese2015}. Thus  each main universality class at depinning corresponds to a specific sandpile model.

The result is   surprising, since  a  simple   $\phi^4$-type theory contains all necessary information  to obtain
 the FRG fixed point of CDWs, 
 a disordered system. It does not directly yield the renormalized 2-point function, or the physics of avalanches. As sketched on Fig.~\ref{f:sector-illustration}, our understanding is that the different field theories     are not equivalent, but when restricted to the same {\em physical sector} make the same predictions. This opens a path to eventually tackle other   systems which necessitate  FRG  via a simpler scalar field theory.

Finally, the   mapping of $\phi^4$-theory at $n=-2$ onto LERWs was done at a microscopic coupling $g=1$.
 Changing the latter
to $p<1$  can be interpreted as a random walk where loops are erased with probability $p$. Since the RG fixed point is reached for any $0<p<1$, we conjecture that these {\em partially} loop-erased random walks  are in the same universality class as LERWs.
We   propose   taking $p$ close to 1  to measure the correction-to-scaling exponent $\omega$ precisely.
While its  $\epsilon$ expansion is known to 6-loop order \cite{KompanietsPanzer2017}, it is only slowly converging, and  we  estimate $\omega = 0.83 \pm 0.01$.

\kaysubsection{Acknowledgements}
It is a pleasure to thank E.~Br\'ezin, J.~Cardy, F.~David,  K.~Gawedzki, P.~Grassberger, J.\ Jacobsen,
 M.V.~Kompaniets, A.~Nahum, S.~Rychkov, D.~Wilson and J.~Zinn-Justin for valuable discussions.

\newcommand{\doi}[2]{\href{http://dx.doi.org/#1}{#2}}
\newcommand{\arxiv}[1]{\href{http://arxiv.org/abs/#1}{#1}}

\appendix



\kaysubsection{Rewriting the operator $\cal O$}
\label{app:calO}
We had defined in \Eq{calO}
\be
{\cal O}(x,y,z) :=    {  \phi}^*_1(x)\Big[{  \phi}_1^*(y) {  \phi}_1 (y)  - {  \phi}_2^*(y) {  \phi}_2 (y)  \Big] {  \phi}_1(z)\ .
\ee
Due to the symmetry of the action under the mapping $\phi_{2}\to -\phi_{2}$, (ibid.\ for $\phi_2^*$) with all other components fixed, it follows that
\bea\label{30}
 0 &=& \left<     \rule{0mm}{2.7mm}
 \phi_{1}^*(x) \phi_{2}(z)   \right> = \left<    {  \phi}_1^*(x)\Big[{  \phi}_1^*(y) {  \phi}_1 (y)  - {  \phi}_2^*(y) {  \phi}_2 (y)  \Big] {  \phi}_2(z)  \right>  \ .~~~~~~~
 \eea
Symmetry upon exchange of components 1 and 2 further yields
\be\label{47}
\left<    {  \phi}_1^*(x)\Big[{  \phi}_1^*(y) {  \phi}_1 (y)  - {  \phi}_2^*(y) {  \phi}_2 (y)  \Big] {  \phi}_1(z) \right> =-  \left<    {  \phi}_2^*(x)\Big[{  \phi}_1^*(y) {  \phi}_1 (y)  - {  \phi}_2^*(y) {  \phi}_2 (y)  \Big] {  \phi}_2(z) \right> \ .
\ee
Using Eqs.~\eq{30} and \eq{47} we deduce that
\bea
&& {\left<   {  \phi}^*_1(x)\Big[{  \phi}_1^*(y) {  \phi}_1 (y)  - {  \phi}_2^*(y) {  \phi}_2 (y)  \Big] {  \phi}_1(z) \right> }
 \nn\\
 && \qquad = \frac12  \left<  [ {  \phi}_1^* (x)-{  \phi}_2^*(x)] \Big[{  \phi}_1^*(y) {  \phi}_1 (y)  - {  \phi}_2^*(y) {  \phi}_2 (y)  \Big] { [ {  \phi}_1(z)+ \phi}_2(z)]\right>\ .
\eea
Define (ibid.\ for the complex conjugate fields)
\be\label{49}
\phi_{+}(x) =\frac1{\sqrt 2} \left[ \phi_{1}(x)+\phi_{2}(x)\right] \ , \qquad
\phi_{-}(x) =\frac1{\sqrt 2}  \left[ \phi_{1}(x)-\phi_{2}(x)\right]\ .
\ee
Then expectations of the observable ${\cal O}(x,y,z)$ can be rewritten as
\bea
\left< \rule{0mm}{2.7mm}{\cal O}(x,y,z) \right>    =  \left< \rule{0mm}{2.7mm}\phi_{-}^*(x)  \phi_{+}(y)\phi_{-}^*(y) \phi_{+}(z)   \right> \ .
\eea
\Eq{49} is  a rotation. Relabeling $\phi_{-}\to \phi_{1}$ and $\phi_{+}\to \phi_{2}$,
this proves the  equivalence of the definitions \eq{calO} and \eq{calObis}.

\kaysubsection{Renormalization group functions and critical exponents} 
The $\beta$-function is defined as
\be
\beta (g) := - m \partial_m g(g_0)\Big|_{g_0=g_0(g)}\ .
\ee
It has a non-trivial fixed point at $g^*$, s.t.\ $\beta(g^*)=0$.
The slope of the $\beta$-function at $g=g^*$ yields the correction-to-scaling exponent $\omega$,
\bea
\omega: =- \beta'(g^*)= \epsilon &-&\frac{2 \epsilon ^2}{3}+\left(\frac{4 \zeta (3)}{3}+\frac{5}{9}\right) \epsilon
   ^3+   \left(\frac{\pi ^4}{90}-\frac{4 \zeta (3)}{3}-\frac{70 \zeta (5)}{9}-\frac{31}{54}\right) \epsilon
   ^4 \nn\\
   &+&     \left( \frac{19 \zeta (3)^2}{27}+\frac{163 \zeta (3)}{54}+\frac{479 \zeta
   (5)}{81}+\frac{833 \zeta (7)}{18}-\frac{5 \pi ^6}{486}-\frac{\pi
   ^4}{90}+\frac{23}{36}\right)  \epsilon ^5 \nn\\ &+& {\cal O} (\epsilon ^6 )
\eea
This result agrees with the literature  on the $O(n)$-model \cite{KleinertNeuSchulte-FrohlindeLarin1991,KompanietsPanzer2017} and for CDWs \cite{HusemannWiese2017} (where $\omega:=\beta'(g^*) $).
To obtain the exponent $z$, we use that
\be
{\left< \int_{{z}} \phi_1^*(x)\phi_1(z) \right>} = \frac 1 {m^2}\ , \qquad
\frac{\left< \int_{{y,z}} {\cal O}(x,y,z) \right>}{\left< \int_{{z}} \phi_1^*(x)\phi_1(z) \right>} \equiv m^2  {\left< \int_{{x}} {\cal O}(x,y,z) \right>} \sim m^{-z}\ .
\ee
With the help of Eqs.~\eq{geff} and \eq{Oeff}, this yields
\be\label{z-from-gamma-tilde}
z = 2  +   \frac{\partial  \gamma_{\phi\phi}}{\partial g} \beta(g)\Big|_{g=g^*} \ .
\ee
The result is given in \Eq{eq:On-24} in the main text. Note that when using the diagrams of Ref.~\cite{KompanietsPanzer2017}, the mass $m$ gets replaced by a momentum scale. Due to universality, the critical exponents are independent of the scheme.

\kaysubsection{The effective coupling at 5-loop order}
\bea
\lefteqn{g =g_0 -3 g_0^2 \bigg[ \textdiagram{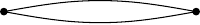}\bigg]  + 3 g_0^3 \bigg[\textdiagram{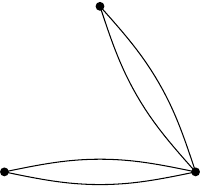} + 4 \textdiagram{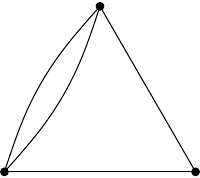}
 \bigg] }\nn\\
 && - 3 g_0^4\bigg[ \textdiagram{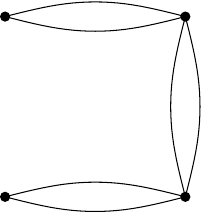}+ 4 \textdiagram{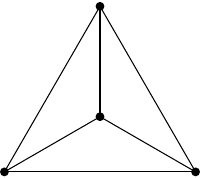}+2  \textdiagram{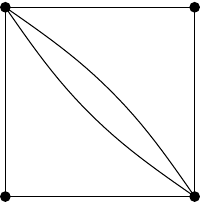} + 16 \textdiagram{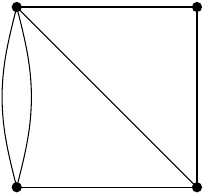} + 4 \textdiagram{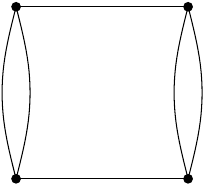} +   4 \textdiagram{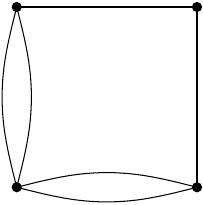} + 4 \textdiagram{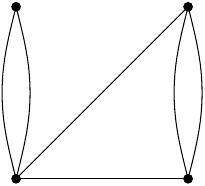}\bigg]\nn  \\
&& +3 g_0^5\Bigg[1\textdiagram{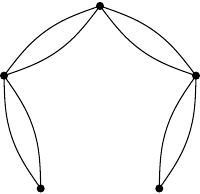}+ 4\textdiagram{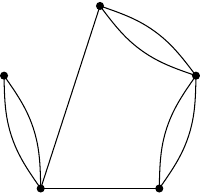}
+ 2\textdiagram{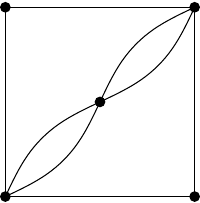}+8\textdiagram{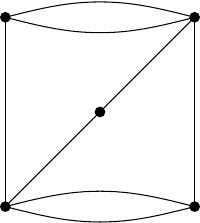}
+4\textdiagram{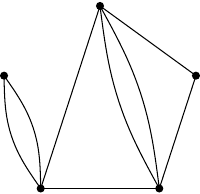}+16\textdiagram{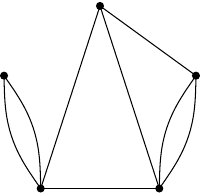}
+4\textdiagram{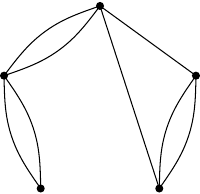}+8\textdiagram{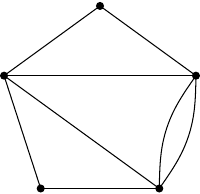}
\nn\\ &&\qquad
+8\textdiagram{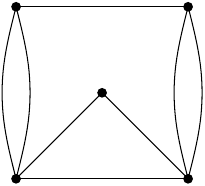}
+16\textdiagram{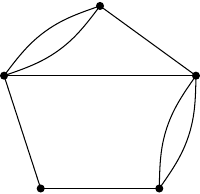}
+8\textdiagram{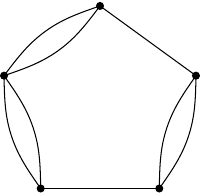}
+24\textdiagram{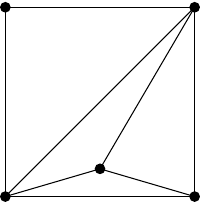}
+32\textdiagram{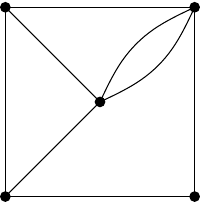}
+32\textdiagram{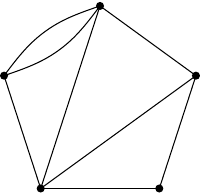}
+16\textdiagram{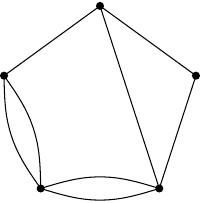}
\nn\\ &&\qquad
+48\textdiagram{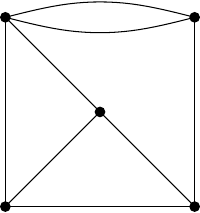}
+8\textdiagram{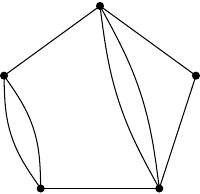}
+32\textdiagram{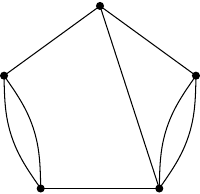}
+4\textdiagram{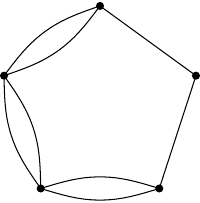}
+8\textdiagram{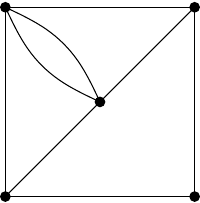}
+4\textdiagram{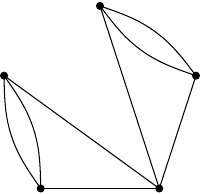}
+28\textdiagram{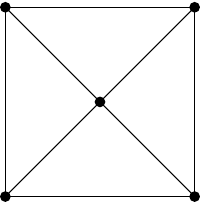}
\Bigg] \nn \\
&& -3 g_0^6\Bigg[
8\textdiagram{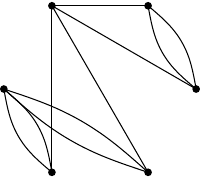}
+32\textdiagram{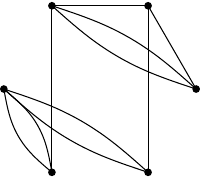}
+16\textdiagram{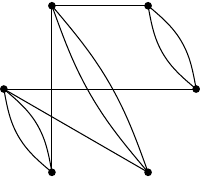}
+112\textdiagram{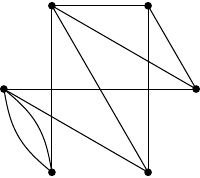}
+96\textdiagram{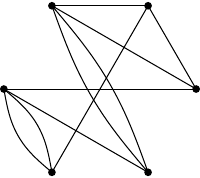}
+4\textdiagram{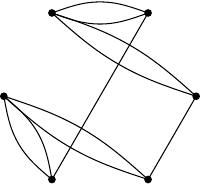}
+4\textdiagram{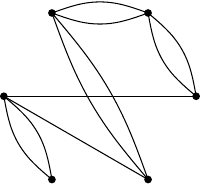}\nn\\&& \qquad
+16\textdiagram{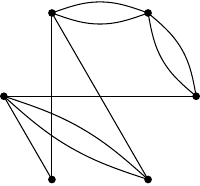}
+24\textdiagram{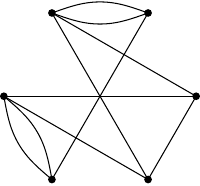}
+32\textdiagram{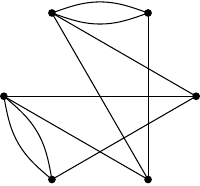}
+8\textdiagram{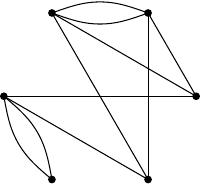}
+32\textdiagram{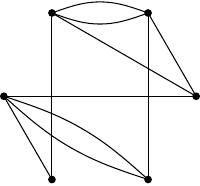}
+4\textdiagram{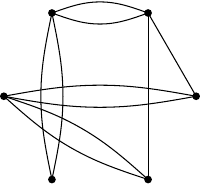}
+8\textdiagram{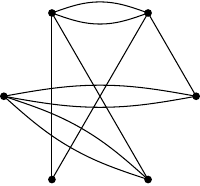}
\nn\\&& \qquad
+1\textdiagram{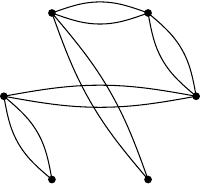}
+2\textdiagram{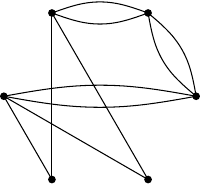}
+4\textdiagram{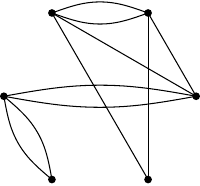}
+8\textdiagram{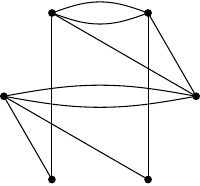}
+16\textdiagram{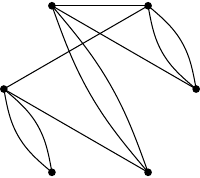}
+32\textdiagram{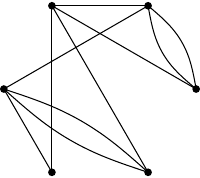}
+32\textdiagram{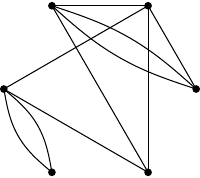}
\nn\\&& \qquad
+16\textdiagram{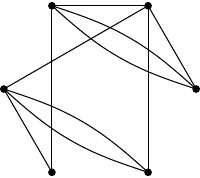}
+32\textdiagram{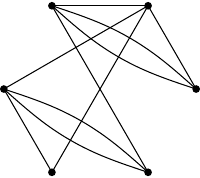}
+36\textdiagram{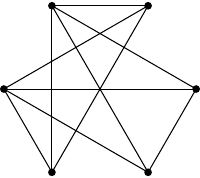}
+56\textdiagram{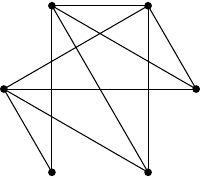}
+96\textdiagram{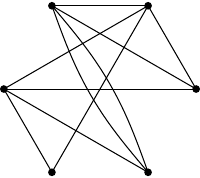}
+2\textdiagram{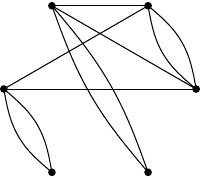}
+4\textdiagram{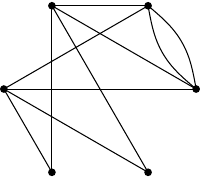}
\nn\\&& \qquad
+8\textdiagram{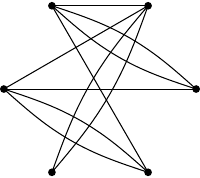}
+16\textdiagram{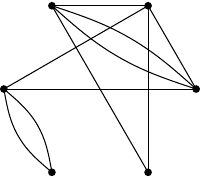}
+32\textdiagram{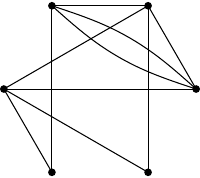}
+4\textdiagram{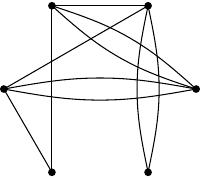}
+8\textdiagram{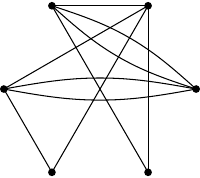}
+32\textdiagram{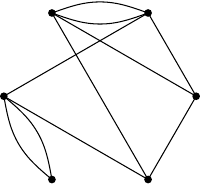}
+32\textdiagram{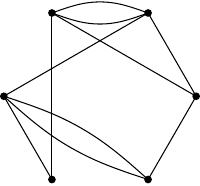}
\nn\\&& \qquad
+32\textdiagram{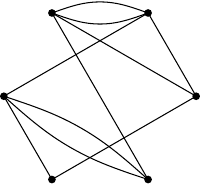}
+16\textdiagram{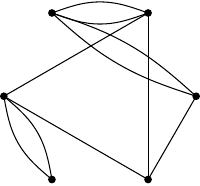}
+16\textdiagram{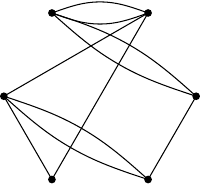}
+16\textdiagram{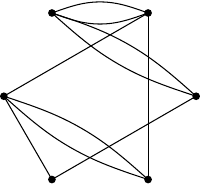}
+96\textdiagram{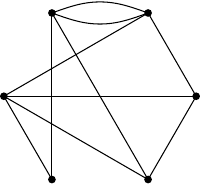}
+32\textdiagram{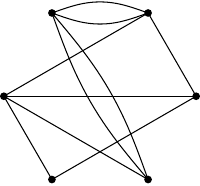}
+16\textdiagram{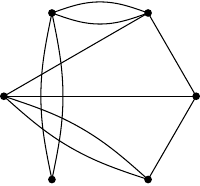}
\nn\\&& \qquad
+32\textdiagram{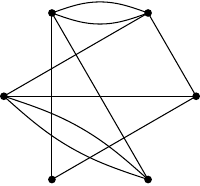}
+48\textdiagram{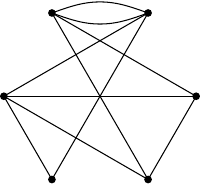}
+64\textdiagram{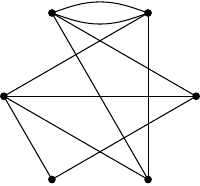}
+32\textdiagram{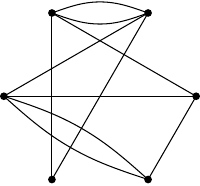}
+32\textdiagram{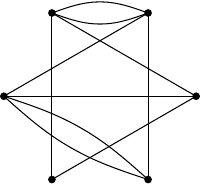}
+32\textdiagram{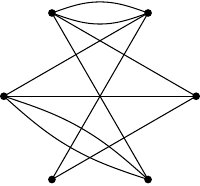}
+32\textdiagram{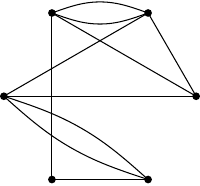}
\nn\\&& \qquad
+16\textdiagram{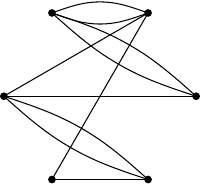}
+4\textdiagram{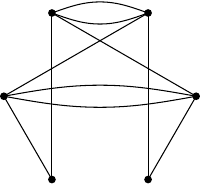}
+4\textdiagram{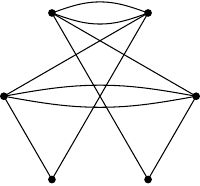}
+4\textdiagram{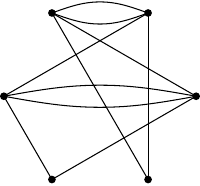}
+4\textdiagram{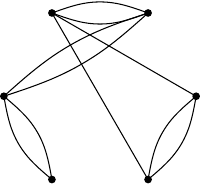}
+8\textdiagram{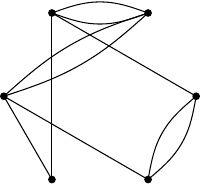}
+32\textdiagram{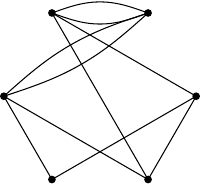}
\nn\\&& \qquad
+16\textdiagram{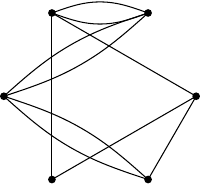}
+4\textdiagram{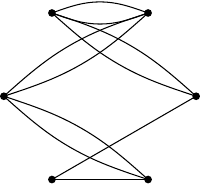}
+8\textdiagram{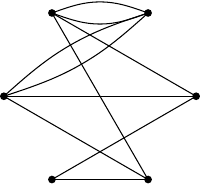}
+64\textdiagram{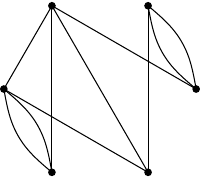}
+32\textdiagram{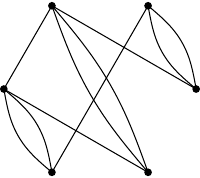}
+48\textdiagram{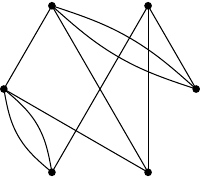}
+48\textdiagram{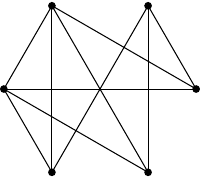}
\nn\\&& \qquad
+96\textdiagram{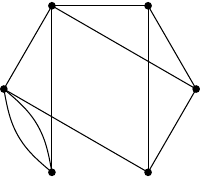}
+224\textdiagram{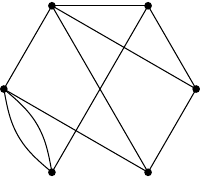}
+32\textdiagram{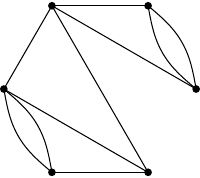}
+48\textdiagram{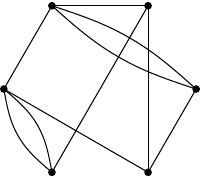}
+16\textdiagram{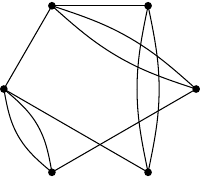}
+32\textdiagram{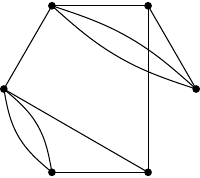}
+272\textdiagram{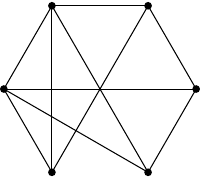}
\nn\\&& \qquad
+48\textdiagram{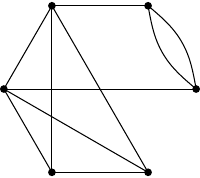}
+8\textdiagram{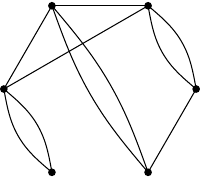}
+48\textdiagram{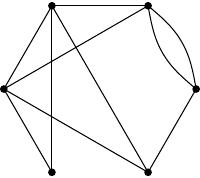}
+32\textdiagram{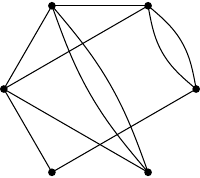}
+24\textdiagram{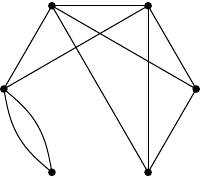}
+112\textdiagram{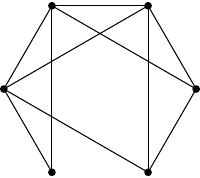}
+96\textdiagram{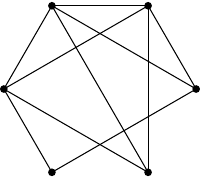}
\nn\\&& \qquad
+64\textdiagram{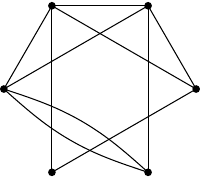}
+12\textdiagram{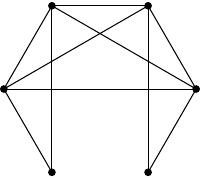}
+8\textdiagram{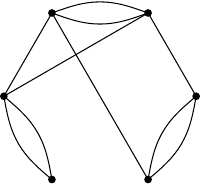}
+32\textdiagram{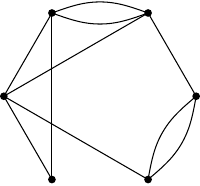}
+64\textdiagram{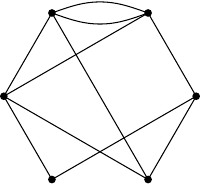}
+32\textdiagram{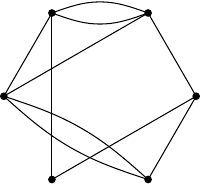}
+32\textdiagram{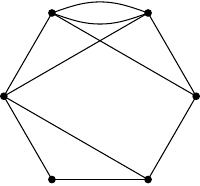}
\nn\\&& \qquad
+16\textdiagram{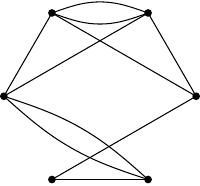}
+8\textdiagram{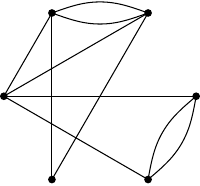}
+64\textdiagram{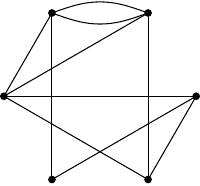}
+32\textdiagram{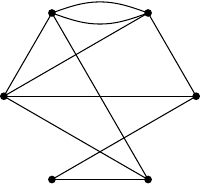}
+8\textdiagram{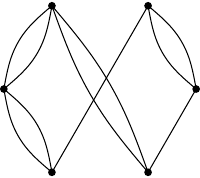}
+16\textdiagram{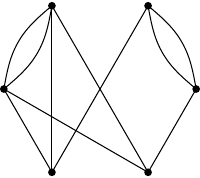}
+48\textdiagram{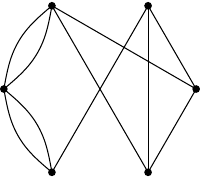}
\nn\\&& \qquad
+32\textdiagram{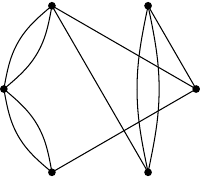}
+96\textdiagram{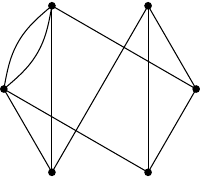}
+64\textdiagram{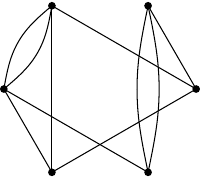}
+8\textdiagram{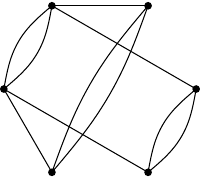}
+28\textdiagram{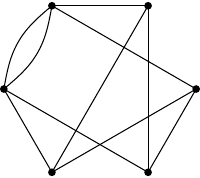}
+56\textdiagram{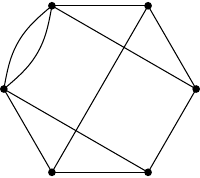}
+4\textdiagram{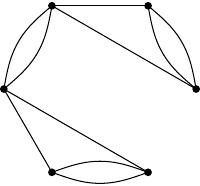}\Bigg]\nn\\
&& + ...
\label{geff}
\eea
Note that at 5-loop order the vertices are sitting on a circle, and crossings inside this circle are not vertices. There are $(2n+1)!!$ terms at $n$-loop order.

\kaysubsection{The observable $\cal O$ at 5-loop order}
Amputating the external points of the observable $\cal O$, and integrating over all but one point, we find
\bea\nn
\lefteqn{\rme^{  \gamma_{\phi\phi}(g_{0})} = m^{4}\left < \int_{{x,z}} {\cal O}(x,y,z) \right>} \\
 &=& 1 -  g_{0}\textdiagram{I1}  + g_{0}^2 \bigg[\textdiagram{3pt2loop1amp} + 2\textdiagram{3pt2loop2amp}
\bigg] \nn\\&&
- g_{0}^{3} \bigg[ \textdiagram{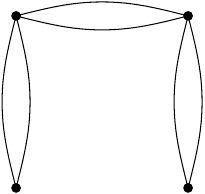} + 2 \textdiagram{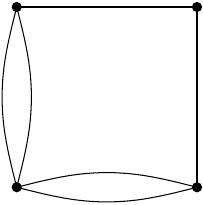}+ 2 \textdiagram{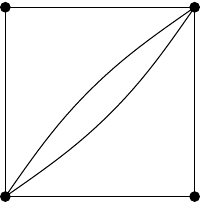}  + 8\textdiagram{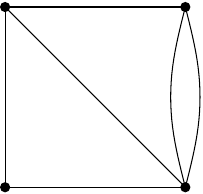}  + 2\textdiagram{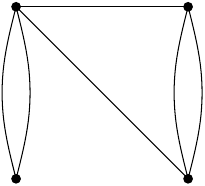}\bigg] \nn
\\
&&  + g_0^4 \bigg[ \textdiagram{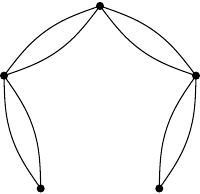} +2 \textdiagram{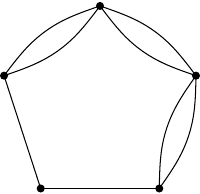}+4 \textdiagram{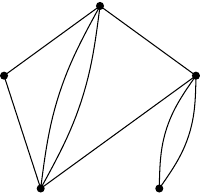} +8\textdiagram{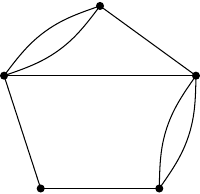}+ 2 \textdiagram{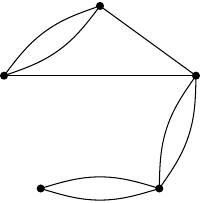}+4   \textdiagram{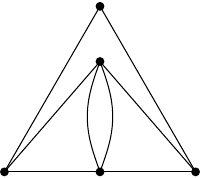}  + 8   \textdiagram{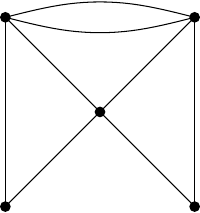}+ 16  \textdiagram{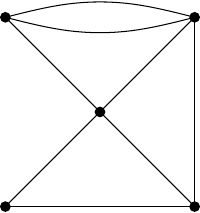} \nn
\\
&& \qquad+16  \textdiagram{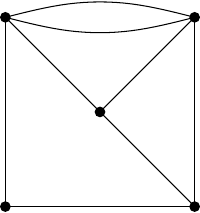} + 8  \textdiagram{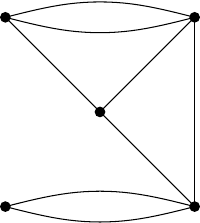}+2   \textdiagram{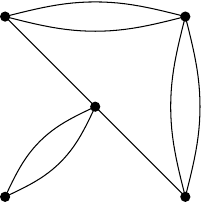}+8  \textdiagram{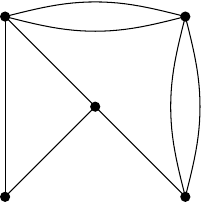}  +2  \textdiagram{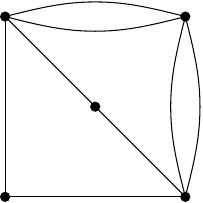}+ 4   \textdiagram{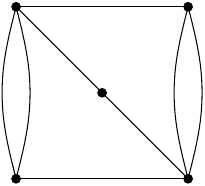}+4   \textdiagram{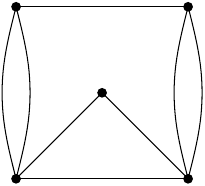}
\nn\\
&& \qquad + 4   \textdiagram{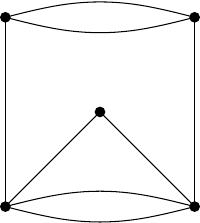}+ 12  \textdiagram{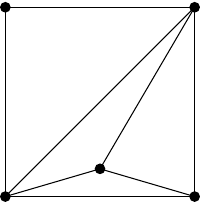}  \bigg] \nn \\
&&  - g_0^5 \Bigg[ \textdiagram{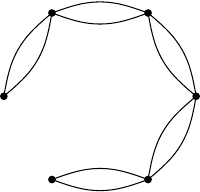}+2\textdiagram{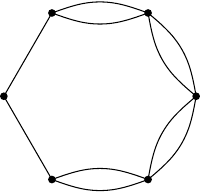}+4\textdiagram{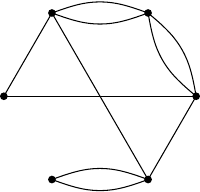} +8\textdiagram{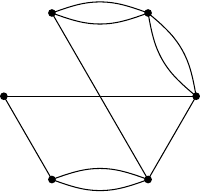}
+2\textdiagram{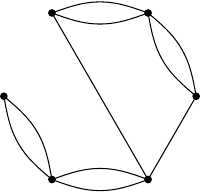}
+4\textdiagram{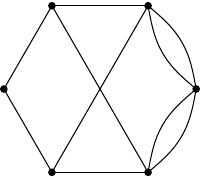}
+4\textdiagram{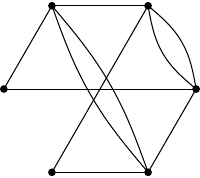}
+16\textdiagram{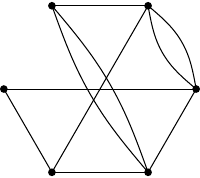}
\nn\\
&& \qquad
+4\textdiagram{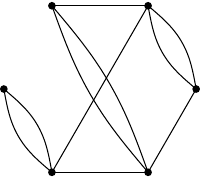}
+2\textdiagram{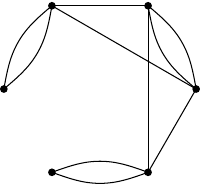}
+8\textdiagram{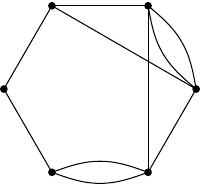}
+4\textdiagram{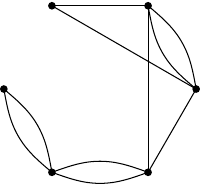}
+16\textdiagram{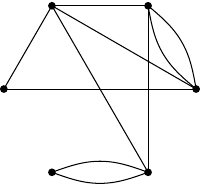}
+16\textdiagram{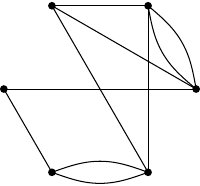}
+16\textdiagram{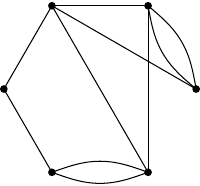}\nn\\
&& \qquad
+8\textdiagram{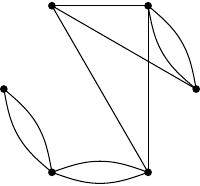}
+8\textdiagram{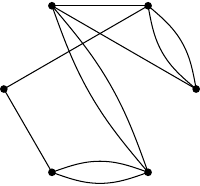}
+2\textdiagram{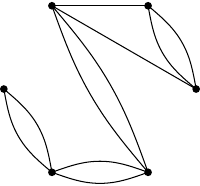}
+16\textdiagram{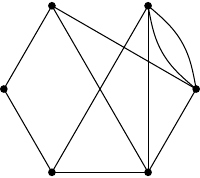}
+4\textdiagram{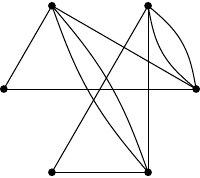}
+16\textdiagram{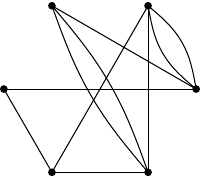}
+4\textdiagram{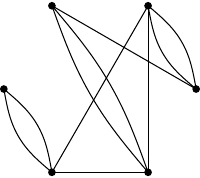}\nn\\
&& \qquad
+8\textdiagram{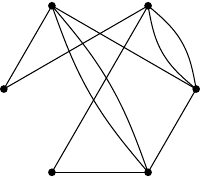}
+16\textdiagram{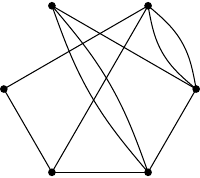}
+16\textdiagram{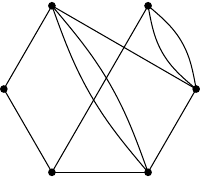}
+8\textdiagram{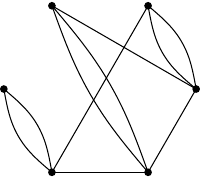}
+4\textdiagram{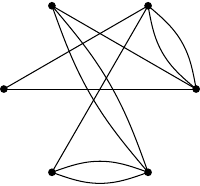}
+8\textdiagram{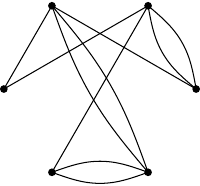}
+8\textdiagram{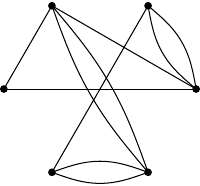}\nn\\
&& \qquad
+4\textdiagram{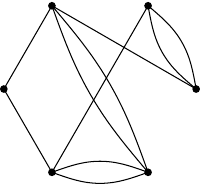}
+12\textdiagram{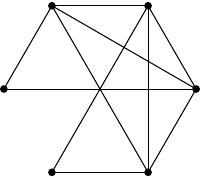}
+48\textdiagram{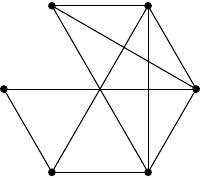}
+12\textdiagram{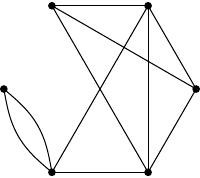}
+32\textdiagram{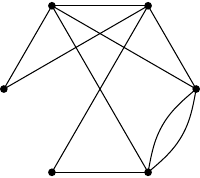}
+32\textdiagram{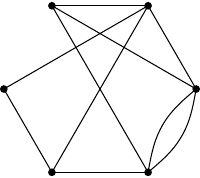}
+32\textdiagram{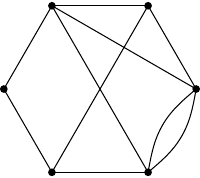}\nn\\
&& \qquad
+16\textdiagram{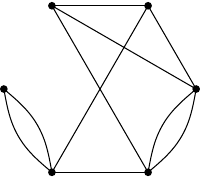}
+32\textdiagram{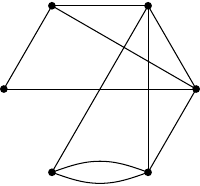}
+16\textdiagram{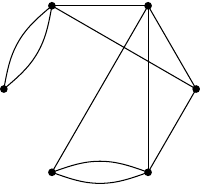}
+32\textdiagram{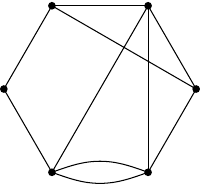}
+16\textdiagram{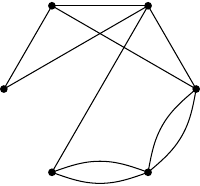}
+8\textdiagram{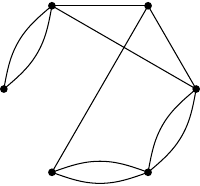}
+8\textdiagram{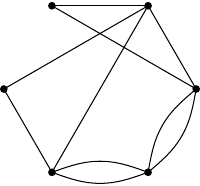}\nn\\
&& \qquad
+16\textdiagram{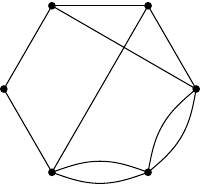}
+48\textdiagram{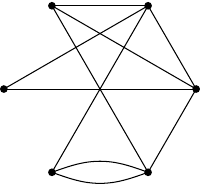}
+24\textdiagram{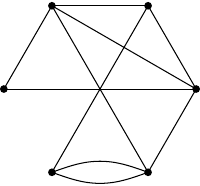}
+24\textdiagram{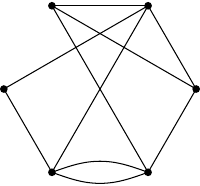}
+48\textdiagram{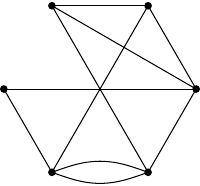}
+16\textdiagram{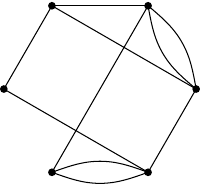}
+4\textdiagram{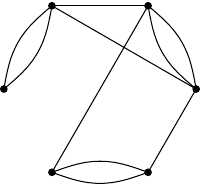}\nn\\
&& \qquad
+4\textdiagram{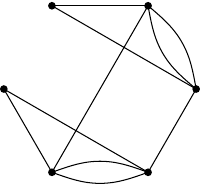}
+16\textdiagram{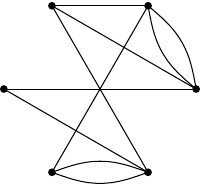}
+16\textdiagram{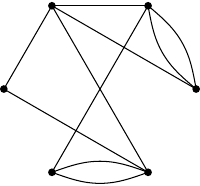}
+16\textdiagram{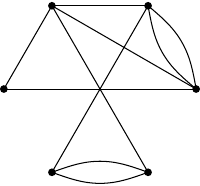}
+16\textdiagram{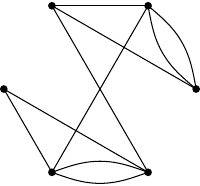}
+16\textdiagram{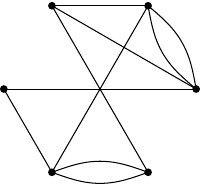}
+16\textdiagram{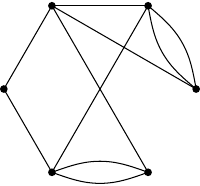}\nn\\
&& \qquad
+8\textdiagram{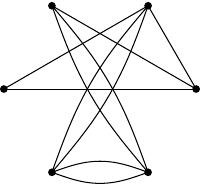}
+2\textdiagram{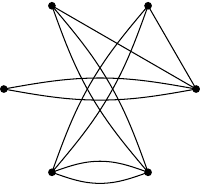}
+2\textdiagram{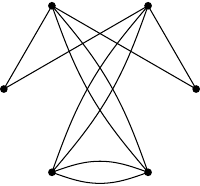}
+4\textdiagram{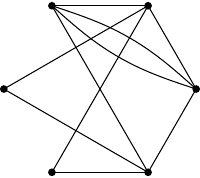}
+16\textdiagram{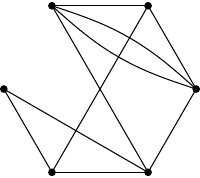}
+4\textdiagram{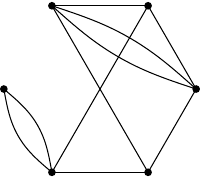}
+8\textdiagram{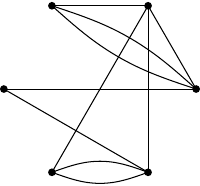}\nn\\
&& \qquad
+4\textdiagram{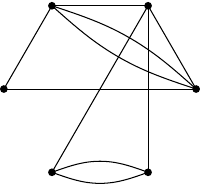}
+28\textdiagram{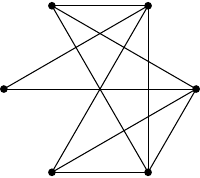}
+56\textdiagram{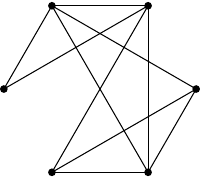} \bigg] + ...
\label{Oeff}
\eea
There are $(2n-1)!!$ diagrams at $n$-loop order. At 5-loop order the vertices are sitting on a circle, and crossings inside this circle are not vertices.


%
%


%
%


\end{document}